\documentclass[pre,aps,twocolumn,superscriptaddress,showpacs,floatfix]{revtex4-1}

\usepackage[english]{babel}
\usepackage{dcolumn}
\usepackage{amssymb}
\usepackage{amsmath}
\usepackage{bm}
\usepackage{color}
\usepackage{graphicx}
\usepackage[usenames,dvipsnames]{xcolor}
\usepackage{balance}

\usepackage{color}
\definecolor{darkblue}{rgb}{0,0,0.6}
\definecolor{darkred}{rgb}{0.6,0,0}

\usepackage{hyperref}
\hypersetup{
colorlinks=true,
urlcolor=darkblue,
citecolor=darkblue,
linkcolor=MidnightBlue,
}

\newcommand{\newtext}[1]{{#1}}
\newcommand{\so}[1]{{}}      % suggests deletion using \so{text}  

\newcommand{\sigmaY}{{\sigma_{\tt Y}}}

\newcommand{\gdot}{{\dot{\gamma}}}

\newcommand{\br}{{\bm r}}

\begin{document}

\title{Criticality in elastoplastic models of amorphous solids with stress-dependent \\ yielding rates}

\author{Ezequiel E. Ferrero} 
\affiliation{Instituto de Nanociencia y Nanotecnolog\'{\i}a, CNEA--CONICET, 
Centro At\'omico Bariloche, (R8402AGP) San Carlos de Bariloche, R\'{\i}o Negro, Argentina.}

\author{Eduardo A. Jagla} 
\affiliation{Centro At\'omico Bariloche, Instituto Balseiro, 
Comisi\'on Nacional de Energ\'ia At\'omica, CNEA, 
CONICET, UNCUYO,\\
Av.~E.~Bustillo 9500
(R8402AGP) San Carlos de Bariloche %R8402AGP S. C. de Bariloche, 
R\'io Negro, Argentina}

%\date{\today}

\begin{abstract}

We analyze the behavior of different elastoplastic models
approaching the yielding transition.
We propose two kind of rules for the local yielding events:
yielding occurs above the local threshold either at a constant
rate or with a rate that increases as the square root of the
stress excess.
We establish a family of ``static'' universal critical exponents
which do not depend on this dynamic detail of the model rules:
in particular, the exponents for the avalanche size distribution $P(S)\sim S^{-\tau_S}f(S/L^{d_f})$
and the exponents describing the density of sites at the verge
of yielding, which we find to be of the form $P(x)\simeq P(0) + x^\theta$
with $P(0)\sim L^{-a}$ controlling the extremal statistics.
On the other hand, we discuss ``dynamical'' exponents that are
sensitive to the local yielding rule details.
We find that, apart form the dynamical exponent $z$ controlling
the duration of avalanches, also the flowcurve's (inverse) Herschel-Bulkley
exponent $\beta$ ($\dot\gamma\sim(\sigma-\sigma_c)^\beta$) enters in this category,
and is seen to differ in $\frac12$ between the two yielding rate cases. 
We give analytical support to this numerical observation by calculating
the exponent variation in the H\'ebraud-Lequeux model and
finding an identical shift.
We further discuss an alternative mean-field approximation to yielding
only based on the so-called Hurst exponent of the accumulated mechanical
noise signal, which gives good predictions for the exponents extracted
from simulations of fully spatial models.
\end{abstract}

\maketitle

%%%%%%%%%%%%%%%%%%%%%%%%%%%%%%%%%%%%%%%%%%%%%%%%%%%%%%%%%%%%%%%%%%%%%%%%%%%%%%%%
\section{Introduction}
\label{sec:intro}

Last years have witnessed major advances in the understanding of the
deformation of amorphous solids from the statistical physics point of
view~\cite{Karmakar2010,lin2014scaling}.
The so-called \textit{yielding transition}, an out-of-equilibrium (driven) transition
between a solid-like and a flowing phase has been analyzed making use of
comparisons and analogies with other well known problems in the literature.
In particular, the depinning transition of elastic manifolds in disordered media~\cite{FisherPR1998}
has largely influenced the interpretation of yielding~\cite{Dahmen2009,tyukodi2015depinning,lin2014scaling}.
In analogy with the classical velocity-force characteristics of depinning $v\sim (f-f_c)^\beta$
(see for example~\cite{FerreroCRP2013}), yielding can be described by a critical behavior
of the steady-state strain rate $\dot \gamma$, that is zero when the stress $\sigma$
is below a critical value $\sigma_c$ and becomes $\dot \gamma\sim(\sigma-\sigma_c)^\beta$
when $\sigma>\sigma_c$. 
Additionally, in the transient regime, the onset of flow %upon deformation
starting from a given glass has been shown to display two qualitatively different
kinds of behavior according to the annealing level, 
%either a discontinuous transition (well annealed glasses) or a
%smooth crossover (poorly annealed glasses), which are 
separated by %an annealing-controlled
a critical point, akin to athermally driven random-field Ising models~\cite{ozawa2018random}. 

In the seek of simplification and generalization, elastoplastic models (EPMs)
built at a coarse-grained level constituted the workhorse in the development 
of a statistical mechanics description of yielding \cite{NicolasRMP2018}.
Nevertheless, the highly phenomenological approach on which these models have
been built supplied a broad variety of rules and details, yielding
different quantitative results and obfuscating the establishment of a set
of universal exponents for the yielding transition.
Despite a broad recent activity on the study of yielding by means of
EPMs, universal properties remain elusive.
Although some consensus has been built in the numerical community
around the avalanche statistics displayed being
different from mean-field depinning~\cite{NicolasRMP2018}, 
quantitatively the reported critical exponents still differ. 
In particular, exponents such as the ones governing the flowcurve 
and the relation between avalanche size and duration show a wayward
behavior.
On the other hand, a comparison of EPMs with well known mean-field constructions
and biased-random-walk problems, lead to a clearer expectation on where one would
find universality and where model details should matter~\cite{jagla2017non}.
More importantly, experiments of yield-stress materials show themselves
a broad variation of exponent values, e.g., $n\simeq 0.2-0.8$ for the
Herschel-Bulkley exponent of the flowcurve~\cite{BonnRMP2017} ($n \equiv \beta^{-1}$).
While this dispersion may be originated in a number of reasons, including a variety
of experimental and measurement protocol details, detailed predictions from a
theoretical perspective, though for an idealized case, could be illuminating.
More than an academic exercise, the forge of consensus around the critical
properties of a problem is crucial in the practical development of the field.
It is with that spirit that we address this issue. % on the modelling side.
%
%analyze here a broad range of models with diverse
%dynamical rules and establish a family of static universal critical exponents.
%At the same time we discuss why dynamical exponents are sensitive to particular system
%details and exemplify their variation, proposing at the same time a
%possible interpretation of typical experimental values for flowcurve exponents.

In this work we contrast and analyze outputs from three different elastoplastic
models previously used in the literature and further modify them to illustrate
other three model cases. 
In particular, we focus on the modification of the rule that governs the onset
of the increase of local plastic deformation, i.e., the local yielding.
In general, when the local stress overcomes a preset threshold ($\sigma_i>\sigma_i^{th}$)
there is a probability $\lambda$ per unit time that a plastic deformation occurs.
In all versions of EPMs presented so far, the value of $\lambda$ is taken as a constant.
However, another alternative seems more natural.
The plastic strain increase when the local stress threshold is overcome can be described
as the passage between a local state that becomes unstable to a new stable state.
%The situation is sketched in Fig. \ref{}.
Therefore, the typical time needed to move to the new minimum, and equivalently the transition
rate $\lambda$, should be a function of the ``degree of instability'' $\sigma_i-\sigma_i^{th}$.
We will refer to this situation as diplaying ``progressive rates'', as opposed to the constant, or uniform case.
When this effect is quantitatively taken into account, the value of the inverse 
Herschel-Bulkley exponent $\beta$ is seen to differ in $\frac12$ between the two cases. 

We make a somewhat {\em ad hoc} classification of exponents governing scaling laws as \textit{static}
and dynamical exponents, relegating to the latter class those exponents
that happen to depend on the particular rules of the model (constant or progressive rates),
as it is the case for $z$ and $\beta$.
Furthermore, we support our numerical observation by estimating analytically
the impact of the ``progressive rates'' modification in the paradigmatic
H\'ebraud-Lequeux model.
Our results highlight an important ingredient to be considered in the
interpretation of the broad range of experimental values observed for the
flowcurve exponent of the yielding phenomenon.

In the athermal and overdamped limit in which the yielding transition is defined,
the quasistatic dynamics is governed by \textit{a priori} uncorrelated collections
of plastic events tagged avalanches.
Events shaping an avalanche, in turn, are supposed to be correlated, giving rise to
a non-trivial dynamics depending on interaction kernel and dimensionality.
Yet, standing on a distant point of the system and assuming ergodicity, all the
physics could be described by the mechanical noise felt by this point due to
the avalanches taking place elsewhere.
By analyzing the time series of the mechanical noise accumulated on time at a
distant location, we can interpret our critical exponents from the problem of
a biased random walk; e.g., giving an explicit expression for $\beta$ in terms
of the Hurst exponent of the accumulated mechanical noise signal, which constitutes
in itself an alternative mean-field proposal for the study of yielding.

%%%%%%%%%%%%%%%%%%%%%%%%%%%%%%%%%%%%%%%%%%%%%%%%%%%%%%%%%%%%%%%%%%%%%%%%%%%%%%%%
\section{Models and simulation protocols}
\label{sec:models}

We consider amorphous materials at a coarse-grained-level description,
laying in between the particle-based simulations and the continuum-level description
provided by classical approaches such as Soft Glass Rheology~\cite{SollichPRE1998}
or the H\`ebraud-Lequeux mean-field model~\cite{Hebraud1998}.
Full background, context and historical development of these so-called
elasto-plastic models (EPMs) can be found in a recent review article~\cite{NicolasRMP2018}.
Briefly, the amorphous solid is represented by a coarse-grained scalar stress
field $\sigma(\br,t)$, at spatial position $\br$ and time $t$ under
an externaly applied shear strain. 
Space is discretized in blocks (e.g., square lattice).
At a given time, each block can be ``inactive'' or ``active''
(i.e., yielding).
This state is defined by the value of an additional variable:
$n(\br,t)=0$ (inactive), or $n(\br,t)=1$ (active).
An over-damped dynamics is imposed for the stress on each block, following some
basic rules:
(i) The stress loads locally in an elastic manner while the block is inactive.
(ii) When the local stress overcomes a local yield stress, a \textit{plastic event} 
occurs with a given probability, and the block becomes ``active'' ($n(\br)$ is set to one).
Upon activation, dissipation occurs locally, and this is expressed as a progressive drop of
the local stress, together with a redistribution of the stresses in the rest of
the system in the form of a long-range elastic perturbation.
A block ceases to be active when a prescribed criterion is met.
The auxiliary binary field $n(\br, t)$ shows up in the equation of motion for the
local stress $\sigma(\br,t)$, defining a dynamics that is typically non-Markovian.
While the structure of the equation of motion for the local stresses is almost unique
in the literature, both its parameters and the rules governing the transitions of
$n(\br)$ ($0\rightleftharpoons 1$) show a variety of choices. %, some of which we describe bellow.

We define our EPMs as a $2$-dimensional scalar field $\sigma(\br,t)$, 
with $\br$ discretized on a square lattice and each block $\sigma_i$ subject
to the following evolution in real space
\begin{equation}\label{eq:eqofmotion1}
\frac{\partial \sigma_i(t)}{\partial t} =
  \mu\dot{\gamma}^{\tt ext}  +\sum_{j} G_{ij} n_j(t)\frac{\sigma_j(t)}{\tau} ;
\end{equation}
where $\dot{\gamma}^{\tt ext}$ is the externally applied strain rate,  and the kernel $G_{ij}$ is the Eshelby
stress propagator~\cite{Picard2004}.

It is sometimes convenient to explicitly separate the $i=j$ term in the previous sum, as
\begin{equation}\label{eq:eqofmotion2}
\frac{\partial \sigma_i(t)}{\partial t} =
  \mu\dot{\gamma}^{\tt ext}  - g_0 n_i(t)\frac{\sigma_i(t)}{\tau} + \sum_{j\neq i} G_{ij} n_j(t)\frac{\sigma_j(t)}{\tau} ;
\end{equation}
where $g_0\equiv -G_{ii} > 0$ (no sum) sets the local stress dissipation rate for an active site.
The form of $G$ is $G(\br,\br') \equiv G(r,\varphi)\sim\frac{1}{\pi r^2}\cos(4\varphi)$ in polar coordinates,
where $\varphi \equiv \arccos((\br-\br')\cdot\br_{\dot{\gamma}^{\tt (ext)}})$ and
$r \equiv \left|\br-\br'\right|$. For our simulations we obtain $G_{ij}$ from the values of the propagator in Fourier space $G_{\bf q}$, defined as
\begin{equation}
G_{\bf q} = -\frac{4q_x^2q_y^2}{(q_x^2+q_y^2)^2}.
\label{eshelby_kernel}
\end{equation}
for $\bf q\ne 0$ and 
\begin{equation}
G_{\bf q=0}=-\kappa
\label{eshelby_kernel_q0}
\end{equation}
with $\kappa$ a numerical constant (see below).
Note that in our square numerical mesh of
size $L\times L$, $q_x^2$, $q_y^2$ must be understood as 
\begin{equation}
q_{x,y}^2\equiv 2-2\cos\left (\frac{\pi m_{x,y}}{L}\right )
\end{equation}
with $m_{x,y}=0,..., L-1$.

The elastic (e.g. shear) modulus $\mu=1$ defines the stress unit, and the mechanical
relaxation time $\tau=1$, the time unit of the problem.
The last term of (\ref{eq:eqofmotion2}) constitutes a \textit{mechanical noise}
acting on $\sigma_i$ due to the instantaneous integrated plastic activity
over all other blocks ($j\neq i$) in the system.
The picture is completed by a dynamical law for the local state variable 
$n_i=\{0,1\}$. %indicating whereas the system at position $i$ is plastically active ($n=1$) or not ($n=0$).
We define hereafter three different rules corresponding to three different models:

\begin{itemize}

\item[1.] \textit{Picard's  model} \cite{Picard2005}
\begin{equation}\label{eq:rulesPicard}
n_i : \begin{cases} 0 \rightarrow 1 & \mbox{at rate~} \tau_{\tt on}^{-1} \mbox{\quad if \quad} \sigma_i >\sigmaY \\
                    0 \leftarrow 1  & \mbox{at rate~} \tau_{\tt off}^{-1} 
      \end{cases}
\end{equation}
\noindent where $\tau_{\tt on}$ and $\tau_{\tt off}$ are parameters and $P(\sigmaY)=\delta(\sigmaY-1)$.

\item[2.] \textit{Lin's model} \cite{lin2014scaling}

\begin{equation}\label{eq:rulesLin}
n_i : \begin{cases} 0 \rightarrow 1 & \mbox{at rate~} \tau_{\tt on}^{-1} \mbox{\quad if \quad} \sigma_i >\sigmaY \\
                    0 \leftarrow 1  & \mbox{instantaneously} %\mbox{at rate~} \tau_{\tt off}^{-1} $\tau_{\tt off}=0^+$
      \end{cases}
\end{equation}
\noindent where $\tau_{\tt on}=1$  and $P(\sigmaY)=\delta(\sigmaY-1)$.
The plastic stress release is instantaneous and the block becomes inactive
immediately after being activated, all $j\neq i$ neighbors receive a kick
that is proportional to the value of the local stress drop and mediated
by the Eshelby propagator.
The local stress drop value is chosen to be $\sigma_i \pm \epsilon$,
with $\sigma_i$ the local stress value just before yielding and $\epsilon$
a uniformly distributed random variable with amplitude $\epsilon_0=0.1$,
to avoid periodic dynamics effects.

\item[3.] \textit{Nicolas' model} \cite{nicolas2014rheology}
\begin{equation}\label{eq:rulesNicolas}
n_i: \begin{cases} 0 \rightarrow 1 & \mbox{instantaneously if \quad} \sigma_i > \sigmaY \\ 
                   0 \leftarrow 1 & \mbox{when } \int dt' | \partial_t\sigma(t')/\mu + \dot{\gamma}^{pl}(t')|\geq\gamma_c
     \end{cases}
\end{equation}
\noindent here $P(\sigmaY)$ is exponentially distributed (as in \cite{liu2015driving}) and the integral over
$dt'$ accounts for the accumulated plastic deformation of the block after the
last local yielding.

\end{itemize}

Besides their differences, one thing that these EPMs
as-found-in-the-literature have in common is that they have a \textit{fixed}
or constant transition rate for plastic activation $\lambda_{\tt fix}=\tau_{\tt on}^{-1}$,
be it finite or arbitrarily large as in Nicolas' model.
As already mentioned, a more natural alternative would be to associate a stress-dependent
typical time with the passage between stable states.
This is seen more convincingly by considering models with continuous local disorder potentials.
This analysis (postponed to Section \ref{sec:rel_to_other_models}) reveals that, for a smooth
form of the effective local confining potential, $\tau_{\tt on} \sim (\sigma_i-\sigma_i^{th})^{-1/2}$.
If we want to maintain an implementation in terms of transition rates,
$\lambda \equiv \tau_{\tt on}^{-1}$ should be then expressed as 
\begin{equation}\label{eq:progressive_rate}
\lambda \sim (\sigma_i-\sigma_i^{th})^{1/2}.
\end{equation}
Notice that a situation of constant rate is recovered 
if the local disorder potential is assumed to produce a jump in the force at the transition point,
a situation that occurs, for instance, if the potential is formed by a concatenation of parabolas.
In this case, the time it takes for the local stress to reach the new minimum 
when $\sigma_i>\sigma_i^{th}$ is roughly constant, independent of the degree
of instability.
We can then associate the prescription of a constant transition rate in previous
EPMs to a local disorder potential formed by the concatenation of parabolic pieces.

Therefore, apart form the ``classical'' implementations of the three models described above,
we further include in our present analysis the same three models but modified with
stress-dependent transition rates of the kind $\tau_{\tt on}^{-1}=\lambda_{\tt prog}=(\sigma_i-\sigmaY_i)^{1/2}$.
Other choices of $\eta$ for $\lambda_{\tt prog} \propto (\sigma_i-\sigmaY_i)^\eta$
could be possible. 
Yet, in the analogy, the disordered potential that would produce something different
from $\eta=1/2$ or $\eta=0$ turns out to be very difficult to justify (see Section \ref{sec:rel_to_other_models}).
Anyhow, for $\eta>0$ the intuitive expectation is that local regions that have
exceeded their stress threshold by larger amounts will take precedence in their
fluidization moment over others.
As we will see, the change from $\eta=0$ to $\eta=1/2$ largely impacts
the rheological properties of EPMs.

\subsection{Finite strain-rate protocol}

Starting from an initial configuration $\{\sigma_i\}$ that observes mechanical stability
the system is evolved according to the equation of motion (\ref{eq:eqofmotion2}) with
an elementary discretized time step $dt = 10^{-2}$ (or smaller).
After each time step integration of $\sigma_i$, the $n_i$ are updated
according to the corresponding rules (\ref{eq:rulesPicard}, \ref{eq:rulesLin}
or \ref{eq:rulesNicolas}, or one of their `progressive rate' equivalents).
The calculation of the convolution between $n_j\sigma_j$ and $G_{ij}$ %in (\ref{eq:eqofmotion1})
is done in Fourier space.
If the ${\bf q}=0$ mode of the propagator is set to zero, 
i.e. $\kappa=0$ (Eq. \ref{eshelby_kernel_q0}), then 
$\sum_{i}G_{ij}=0$, and the dynamics is stress conserving.
However, in order to be able to control the strain rate $\dot{\gamma}^{\tt ext}$ we are interested to work with non-conserved stress protocols.
This is accomplished by taking $\kappa>0$. 
By spatially averaging Eq. \ref{eq:eqofmotion1} % (indicating spatial averaging by the overline):
we get
\begin{equation}\label{eq:eq_avg}
\frac{\partial \overline \sigma(t)}{\partial t} =
  \mu\dot{\gamma}^{\tt ext}  -\frac{\kappa}{\tau}  \overline {n\sigma}(t);
\end{equation}
that clearly indicates how $\kappa$ produces a stress decay in the system as
long as there are active sites ($n\ne 0$).
We use $\kappa=1$, as in previous strain-controlled EPMs implementations~\cite{Martens2012,nicolas2014rheology,liu2015driving},
unless otherwise specified.
% which results in $g_0 \simeq 0.57$ in (\ref{eq:eqofmotion}).

It has to be emphasized that initial conditions play no role in our study,
as all measurements are done in the steady state, after a very long straining stage
where the memory of the initial condition is totally lost.
The transient where we don't take any measurements is sufficiently long so that every
block has yielded several times already.
For the flowcurves construction each fix strain-rate has its own long transient strain period.

\subsection{Quasistatic protocol}

For the analysis of avalanche statistics, it is convenient to have a
protocol that allows for the triggering and unperturbed evolution
(no driving) of avalanches until they stop (what is  guaranteed by
a degree of stress non-conservation $\kappa>0$).
This is the quasi-static protocol described here.

Starting from any stable configuration, i.e., no
site is active and no site stress is above its local
threshold ($n_i=0$ and $\sigma_i<\sigmaY_i$ for all sites),
the next avalanche of plastic activity is triggered by
globally increasing the stress by the minimum amount necessary
for a site to reach its local threshold.
That site (the weakest) is activated at threshold with no stochastic delays;
it perturbs the stress values of other sites and the rest of the avalanche evolves
without any external drive following the dynamics prescribed
by Eq. (\ref{eq:eqofmotion2}) (and the corresponding activation rule)
with $\gdot=0$. % (strictly zero).
The avalanche stops once there are no more active sites and all stresses
are below their corresponding thresholds again.
At this point the loading process is repeated.
For each simulation run, data is collected only in the steady-state.

We have noticed that Nicolas' model as originally proposed~\cite{nicolas2014rheology}
is ill-defined in the quasistatic limit.
Eventually one arrives to the pathological case in which a site is active
and the criterion to recover local elasticity (\ref{eq:rulesNicolas}) is never
met (any finite $\gdot$ guarantees it, but not the quasistatic protocol).
To overcome this issue, we add a maximum duration bound to plastic events,
yet large enough to guarantee the full relaxation of the local stress.

%%%%%%%%%%%%%%%%%%%%%%%%%%%%%%%%%%%%%%%%%%%%%%%%%%%%%%%%%%%%%%%%%%%%%%%%%%%%%%%%
\section{Results}
\label{sec:results}

\subsection{Quasistatic avalanche distributions}

\begin{figure}[!tb]
\begin{center}
\includegraphics[width=1\columnwidth]{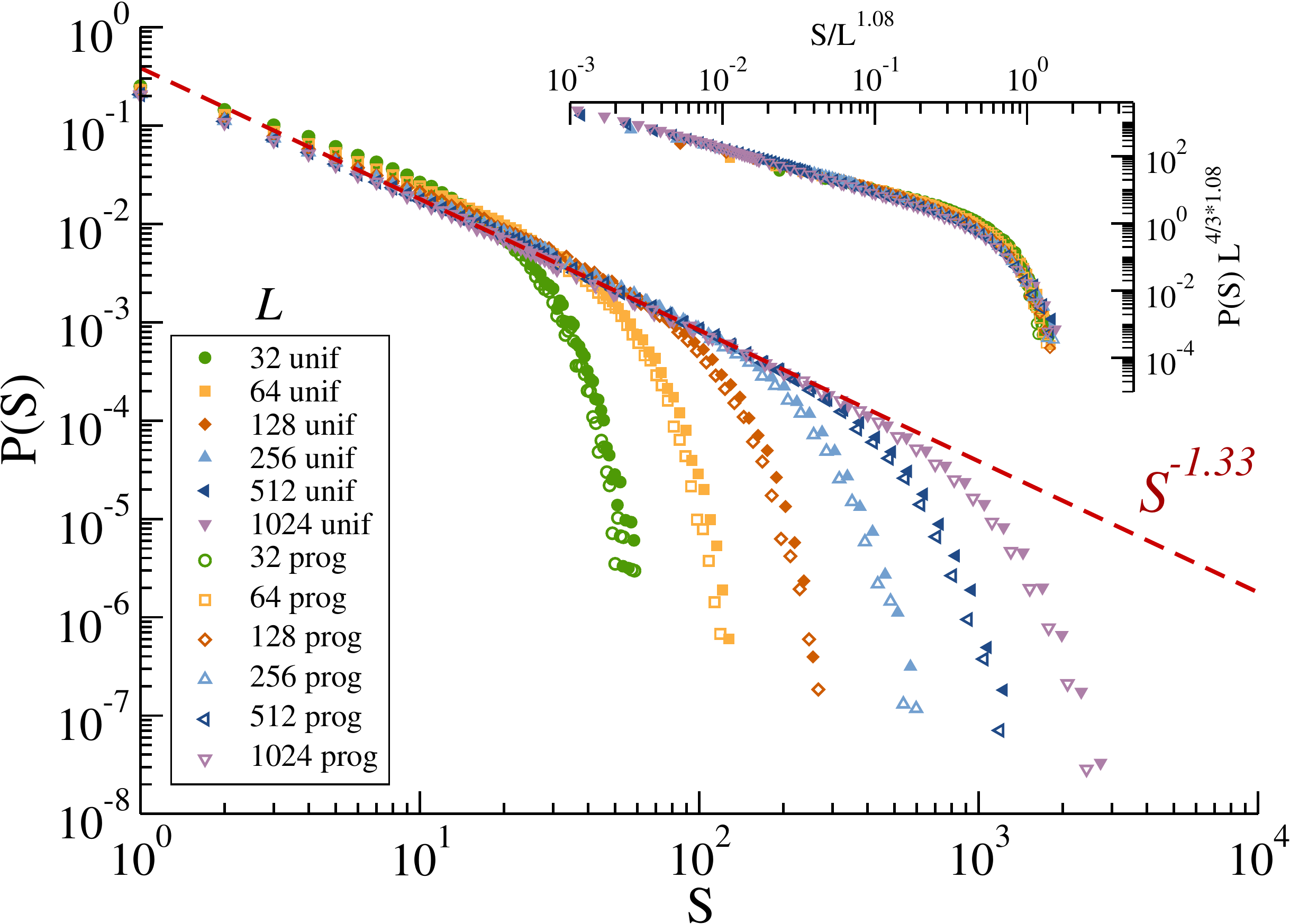} 
\includegraphics[width=1\columnwidth]{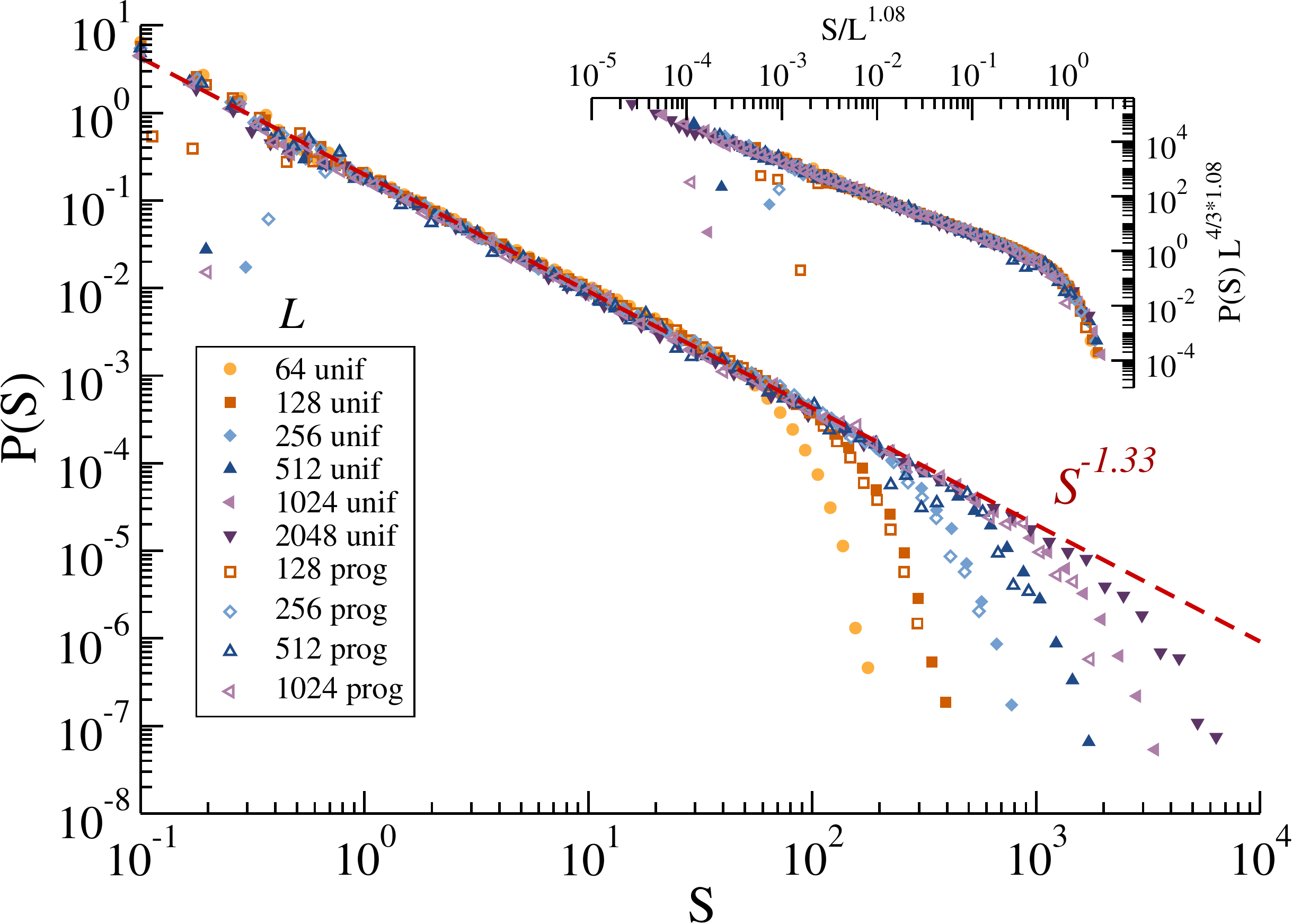} 
\includegraphics[width=1\columnwidth]{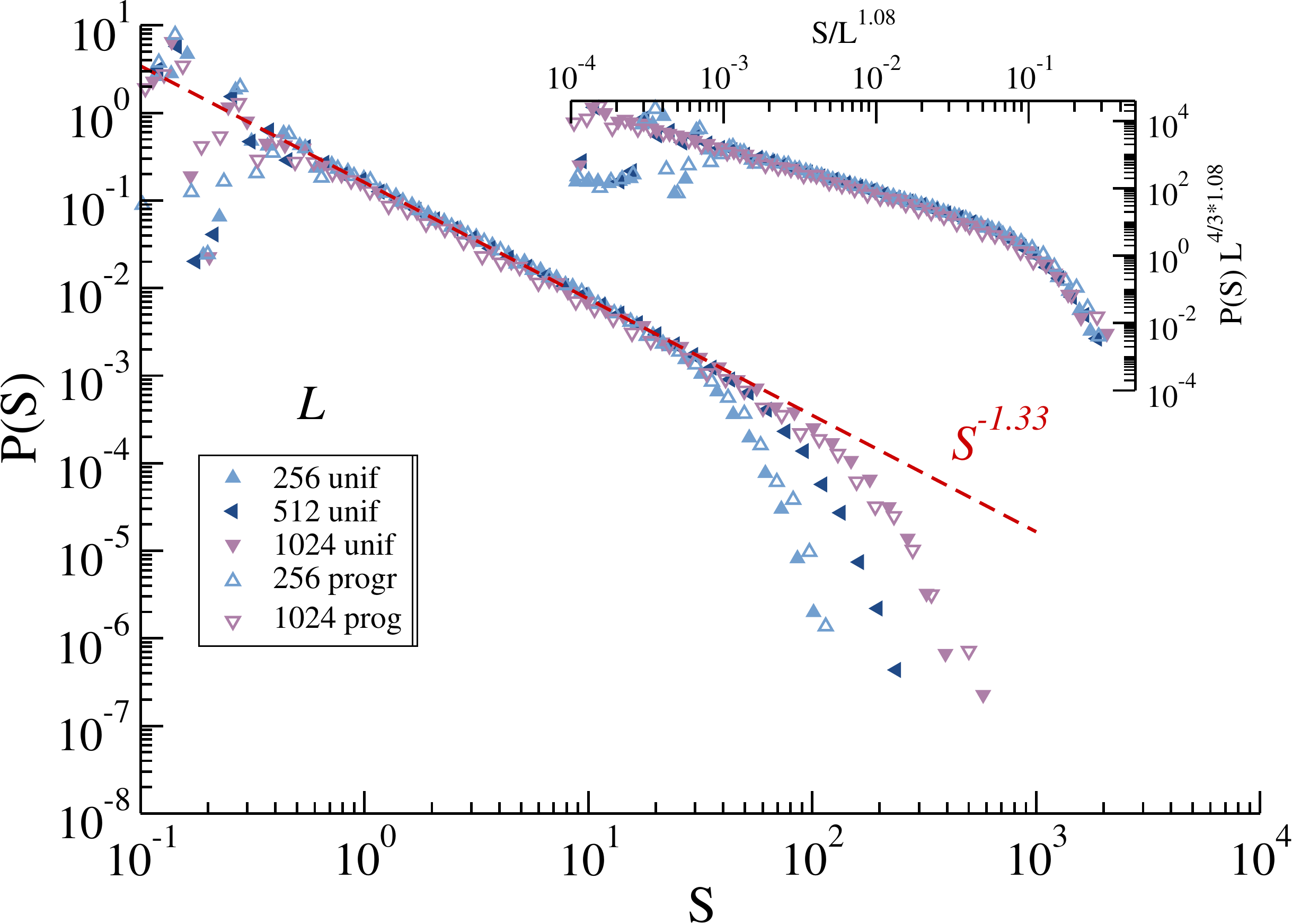}
\end{center}
\caption{
Avalanche size distribution of the quasistatic dynamics for different system sizes
as described by the legends.
\textit{Upper panel:} Lin's model.
\textit{Middle panel:} Picard's model.
\textit{Lower panel:} Nicolas' model.
For each model, results of simulations corresponding to uniform (filled symbols)
and progressive (open symbols) yielding rates are shown.
On each panel, the dashed line displays a power-law $\sim S^{-\tau_S}$ with $\tau_S \simeq 1.33$ 
Insets show the scaling $P(S)L^{\tau_S d_f}$ vs $S/L^{d_f}$, with $d_f\simeq 1.08$,
$\tau_S \simeq 1.33$.
}\label{fig:PofS}
\end{figure}

Starting from a system that has been deformed at a slow strain rate,
we apply a quasistatic protocol.
Size $S$ and duration $T$ of each triggered avalanche are measured.
$S$ is simply calculated from the total stress drop $\Delta\sigma$ caused by the avalanche
of plastic events as $S=\Delta\sigma L^d$, while $T$ is the time elapsed until
the avalanche ceases its activity, measured in units of $dt$.

Figure~\ref{fig:PofS} shows avalanche size distributions
for all three models in their two transition rate variants,
at various system sizes.
All distributions have the form $P(S) \sim S^{-\tau_S}f(S/S_{\tt max})$,
with $f(x)$ a rapidly decaying function (a compressed exponential).
Taking a single EPM %(e.g., Picard's model, fig.\ref{fig:PofS}b)
no difference between uniform and progressive rates is observed.
Furthermore, the same value of $\tau_S\simeq 1.33$ characterizes
all avalanche distributions, irrespective of the six model variants analyzed.
Good collapse of the distributions for different system sizes 
is found when plotting $P(S)L^{\tau_S d_f}$ vs $S/L^{d_f}$ with the
so-called ``fractal dimension'' $d_f=1.08$.
However, explicit and acceptable fittings of
$P(S) = A S^{-\tau_S}\exp(-(S/S_{\tt max})^b)$ with $b \simeq 1.5 \ldots 2$
may result in $S_{\tt max}\propto L^{d_f}$ with $d_f \simeq 1.1 \ldots 1.5$
as well.
In any case, we notice that $d_f$ is independent on the particular form of
the rates used.

\begin{figure}[!ht]
\begin{center}
\includegraphics[width=1\columnwidth]{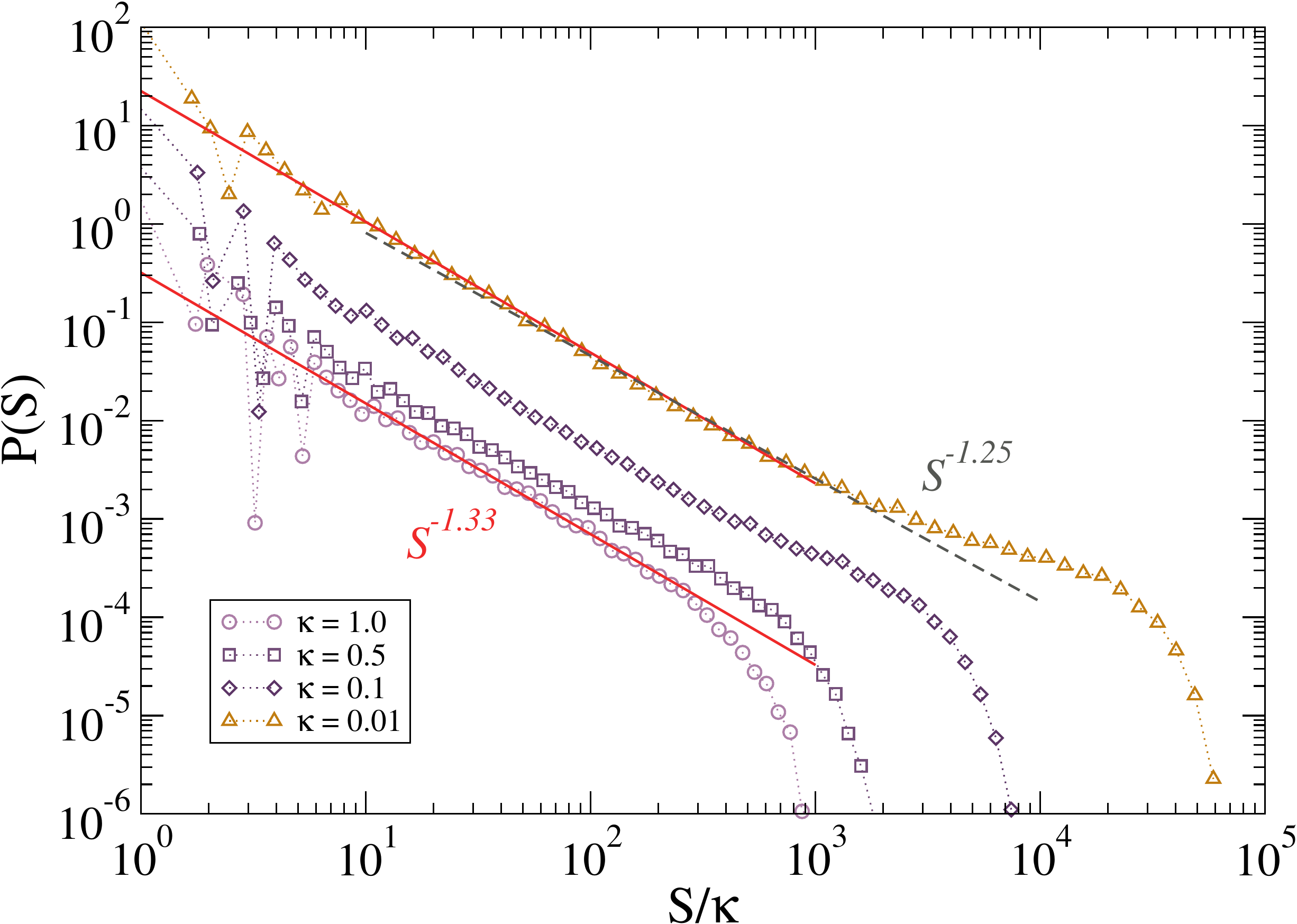} %-crop.pdf} 
\end{center}
\caption{
Avalanche size distribution of the quasistatic dynamics of Lin's model with uniform rates
for system size $N=512^2$ and different choices of the stress non-conservation parameter $\kappa$.
}\label{fig:PofSvariyingGoo}
\end{figure}

Apart from the uncertaintes in the determination of $d_f$, a word should
be said about the accurate quantitative estimation of $\tau_S$.
We have noticed that its estimation can be sensitive to some technicalities.
The avalanche sizes distribution shape changes with the parameter of
stress (non-)conservation $\kappa$ (the value chosen for $-G_{{\bf q}=0}$).
When $\kappa=0$, stress is conserved by the dynamics.
In that limit, we will soon or late observe a never ending
avalanche once the loading takes the system above the critical threshold.
For small but finite values of $\kappa$ any stress excess eventually decays,
but large avalanches are difficult to stop and this results in a plateau or
bump in the $P(S)$ distribution at large $S$.
For relatively small systems sizes, as the ones one can usually simulate, 
an important part of the distribution is impacted by this protocol detail (the value of $\kappa$).
Fig. \ref{fig:PofSvariyingGoo} illustrates (for Lin's model with uniform rates)
how the choice of $\kappa$ affects the quantitative estimation of $\tau_S$ in a typical
finite-size avalanche size distribution. 
If one chooses a power-law window in $P(S)$ neglecting very small and very large $S$
(usually associated with finite time-step $dt$ discretization and finite system size $N$, respectively),
we would observe $\tau_S$ varying from $\tau_S\simeq 1.33$ for $\kappa=1$ to
$\tau_S\simeq 1.25$ for $\kappa = 0.01$.
Still, $\tau_S$ should be independent of the precise value of $\kappa$ in the thermodynamic limit.
%
%In fact, we observe that $\tau_S$ and $d_f$ appear to be quite
%independent of the particular model once a simulation protocol is defined.
%
The particular value of $\kappa$ has an effect on $P(S)$ in the range of $S$
for which the last term in Eq. (\ref{eq:eq_avg}) gives a total stress drop
comparable with the present value of the stress at the yielding sites times $\kappa$.
For an avalanche of size $S$ in a system of size $N\equiv L^d$ 
this occurs when $\Delta\sigma L^{d} \sim \sigmaY L^{d_f} \kappa$ or larger.
This means that for any $\kappa$ we should have a consistent value of $\tau_S$
as long as we fit $P(S)$ for an avalanche range not reaching this limit.
For the system in Fig. \ref{fig:PofSvariyingGoo}, this is $S/\kappa \sim 512^{d_f}$.
On the left of such a value, all curves seems consistent with the power-law found
in Fig.\ref{fig:PofS} (red lines).
This is to argue that, although different exponent values are found in the
literature~\cite{Talamali2011,budrikis2015universality,liu2015driving,lin2014scaling,NicolasRMP2018}
and we cannot be conclusive about $\tau_S\simeq 1.33$ being the universal value,
the sensitivity of $P(S)$ to protocols with effectively different
values of $\kappa$ may justify the spread of reported values of $\tau$. %values in the literature.
Systematic finite-size analysis are needed to reach an agreement
on a unique universal exponent. 
We have done a first step here by clearly demonstrating that, at least when
fixing $\kappa$, $\tau_S$ and $d_f$ do not depend on the yielding rate rule,
nor in other details defining the particular EP model. % in the thermodynamic limit.

\subsection{Extremal statistics and density of shear transformations $P(x)$}
\label{sec:xmin_and_Pofx}

Directly related to the avalanche size statistics during quasistatic simulations
is the stress $x_{\tt min}$ needed to trigger the next avalanche each time
($x_{\tt min}=\sigmaY_w-\sigma_w$, where $w$ stands for the ``weakest'' site).
It was first observed by molecular-dynamics simulations \cite{Maloney2004}
and then contextualized in a framework of extreme statistics both in 
MD \cite{Karmakar2010} and EPMs \cite{lin2014scaling}, that the
finite size scaling of the average loading stress needed to trigger avalanches
is sub-extensive, i.e.,
\begin{equation}\label{eq:xmin}
\left<x_{\tt min}\right> \propto N^{-\phi} ~,~ \text{ with } ~ 1<\phi<2
\end{equation}
with $N=L^d$.
This is at odds with the intuition that we get from analogies with
sand-pile models or other stick-slip dynamics as the one related to the depinning transition.
There, the intensive local variables (pile height or force at site $i$, equivalent
to the stresses $\sigma_i$) can only increase as the intra-avalanche dynamics proceeds and
therefore a larger system will always show an equally weaker site on average ($\phi=1$).

\begin{figure}[!t]
\begin{center}
\includegraphics[width=0.9\columnwidth]{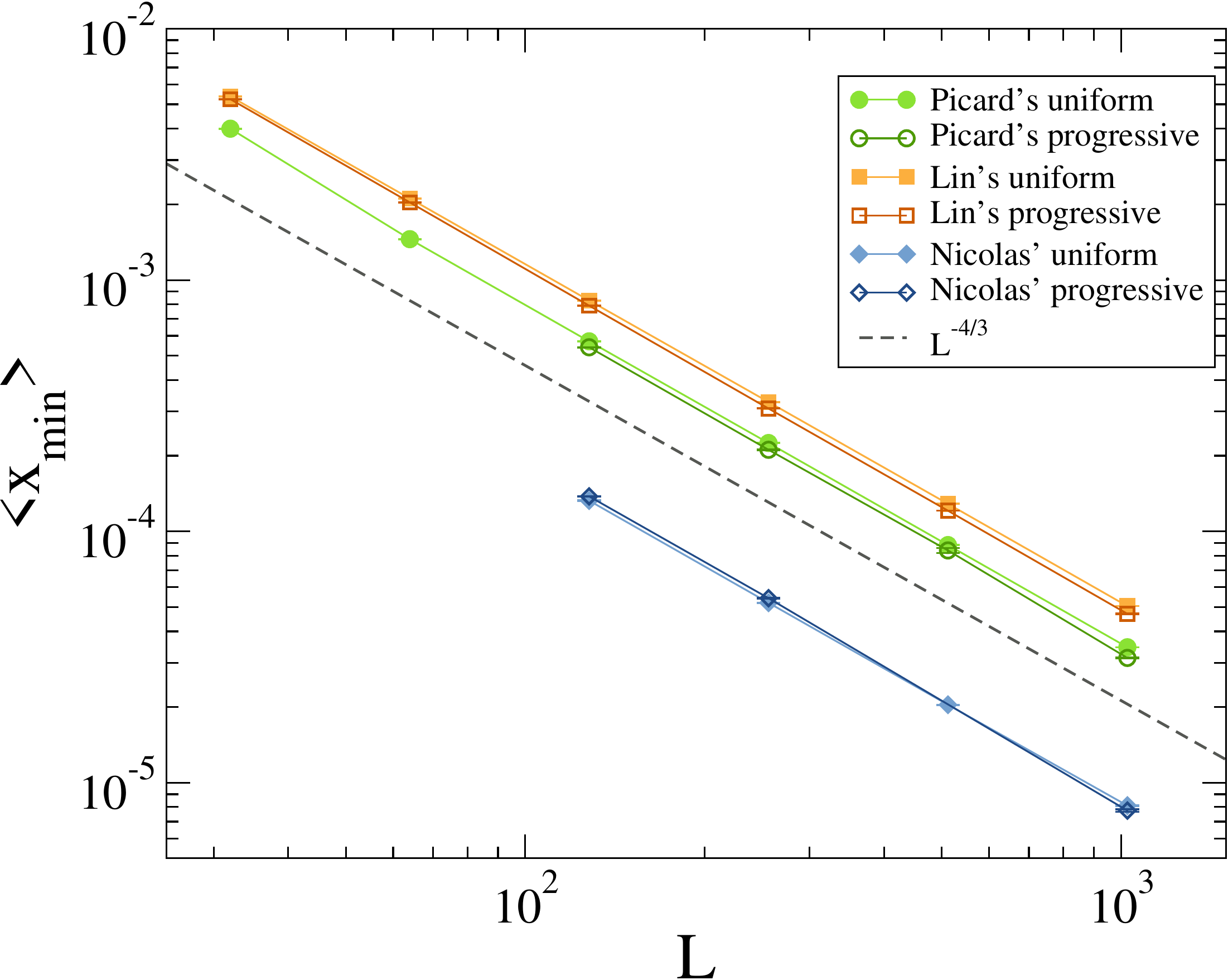} 
\end{center}
\caption{
Finite size scaling of the average stress needed to trigger a new avalanche 
$x_{\tt min}$ for all model variants, as indicated by the legend.
The dashed line shows $\sim L^{-\frac43}$ for comparison.
}\label{fig:xminscaling}
\end{figure}

This phenomenological sub-extensiveness was interpreted as a consequence
of marginal stability in the driven amorphous solid by Lin \textit{et al.}\cite{lin2014scaling}.
In average, starting at any given steady state configuration, sites far away from their
thresholds have a much larger life time before yielding than sites that are close to it.
Since plastic activity provides signed kicks to each site stress, the problem of
local yielding is that of a first passage time, and the probability of finding
the walker away from the boundary is much larger than finding it very close to it.
Analyzing the distribution $P(x)$ of local stresses needed to reach local thresholds
($x_i=\sigmaY_i-\sigma_i$)--tagged the ``density of shear transformations''--and 
assuming it to be characterized by a power-law $P(x)\sim x^\theta$ at
vanishing $x$, it can be shown that the distribution of the minimal 
values ($x_{\tt min}$) among randomly chosen sets of $N\gg 1$
independent samples from $P(x)$ is a Weibull distribution with mean value 
\begin{equation}\label{eq:xminfss}
\left<x_{\tt min}\right> \sim N^{-1/(1+\theta)} = L^{-d/(1+\theta)}.
\end{equation}
In this way, any $\theta>0$ provides a sub-extensive $\left<x_{\tt min}\right>$,
and in fact, $\theta > 0$ was estimated~\cite{Karmakar2010,lin2014scaling,NicolasRMP2018}
for the yielding of amorphous solids in contrast with the $\theta=0$ value
of the depinning of elastic manifolds.
Fig. \ref{fig:xminscaling} shows $\left<x_{\tt min}\right>$ as a function of 
system size for different elastoplastic models obtained in the steady state of
quasistatic simulations.
We can observe that results for the uniform and progressive rates 
variants of the models are totally consistent.
Furthermore, all the simulated EPMs display the same scaling law
$<x_{\tt min}> \sim L^{-1.33}$, compatible with $\theta\simeq 0.5$ in Eq. (\ref{eq:xminfss}).

\begin{figure}[!t]
\begin{center}
\includegraphics[width=1\columnwidth]{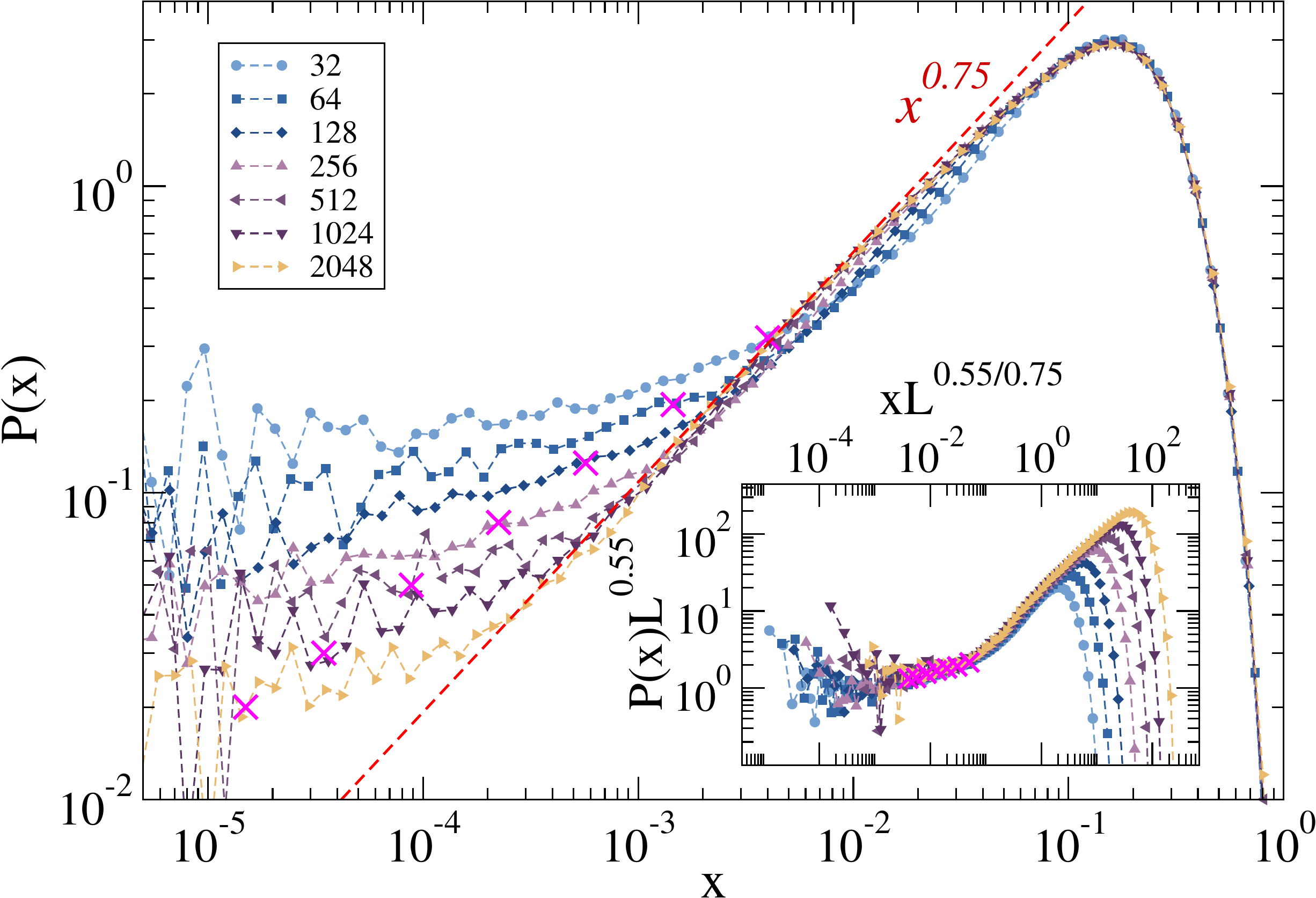} 
\end{center}
\caption{
$P(x)$ for different system sizes as indicated in the labels corresponding
to  Picard's model with uniform rates.
Magenta crosses indicate the position of $\langle x_{\tt min\rangle}$
for each system size.
The inset shows the scaling $P(x)/L^{-a}$ vs. $x/L^{a/\theta}$, with $a=0.55$.
}\label{fig:PofXfinitesize}
\end{figure}

At this point, it is appropriate to clarify a dichotomy in the determination of
exponent $\theta$. 
It turns out that $P(x)$ does not keep a form $P(x)\sim x^\theta$ when
$x\to 0$ as it is commonly assumed.
Instead, it deviates from such power-law.
In Fig. \ref{fig:PofXfinitesize} we show the results obtained for
$P(x)$, as determined from quasistatic simulations collecting 
data right after avalanches have finished and before a new
loading phase begins, in the steady state.
%
%$P(x)$ preserves the excess of probability at $x=0$ even when the $x$ values are
%rigorously collected just after the end of an avalanche and before the next loading stage.
%
The deviation from a common power law, and the establishment of a system-size
dependent plateau at $x\to 0$ is clear, as it is also the fact that this
plateau occurs systematically at smaller values of $x$ as $L$ increases.
Note that this effect can be easily overseen if an arbitrary
lower cutoff of the histograms is imposed.

Therefore, the form of the distribution of stress distances to local threshold
is rather $P(x) \simeq P(0)+x^{\theta}$. 
The emergence of this plateau at small $x$ disentangles $\theta$ from the exponent
$\phi$ that rules the finite size scaling of $\langle x_{\tt min} \rangle$ in Eq.\ref{eq:xmin}.
So we give a new name to the `apparent' $\theta$ holding Eq. \ref{eq:xminfss}:
$\vartheta = d/\phi - 1$, while keeping $\theta$ for a behavior $P(x) \sim x^{\theta}$
in the intermediate range of $x$ where it holds.
The inset of Fig. \ref{fig:PofXfinitesize} shows an empirical scaling
of the form $P(x)=P(0)+x^{\theta}$ with $P(0) \sim L^{-a}$. 
We obtain $a \simeq 0.55$.
$\theta$ is in the range $0.6-0.77$ depending on the criterion
chosen to identify the power-law regime.
Such power-law range is usually limited by finite size effects.
In our case, only for the biggest $L$ simulated we can say that $P(x)$
displays a power-law in more than a decade, with exponent $\theta \simeq 0.75$.
%
%The way $P(0)$ scales with $L$ should be such that $P(0) \left<x_{\tt min}\right> \sim L^{-d}$.
%If $P(0) \sim L^{-a}$ and $\left<x_{\tt min}\right> \sim L^{-b}$, then $a+b=d$ should hold.
%From Fig.\ref{fig:xmin} we know $b \simeq 1.34$, so $a$ should be $\sim 0.66$.
%A value of $a\simeq 0.55$ gives a better scaling in Fig.\ref{fig:PofXfinitesize} though.

\begin{figure}[!t]
\begin{center}
\includegraphics[width=1\columnwidth]{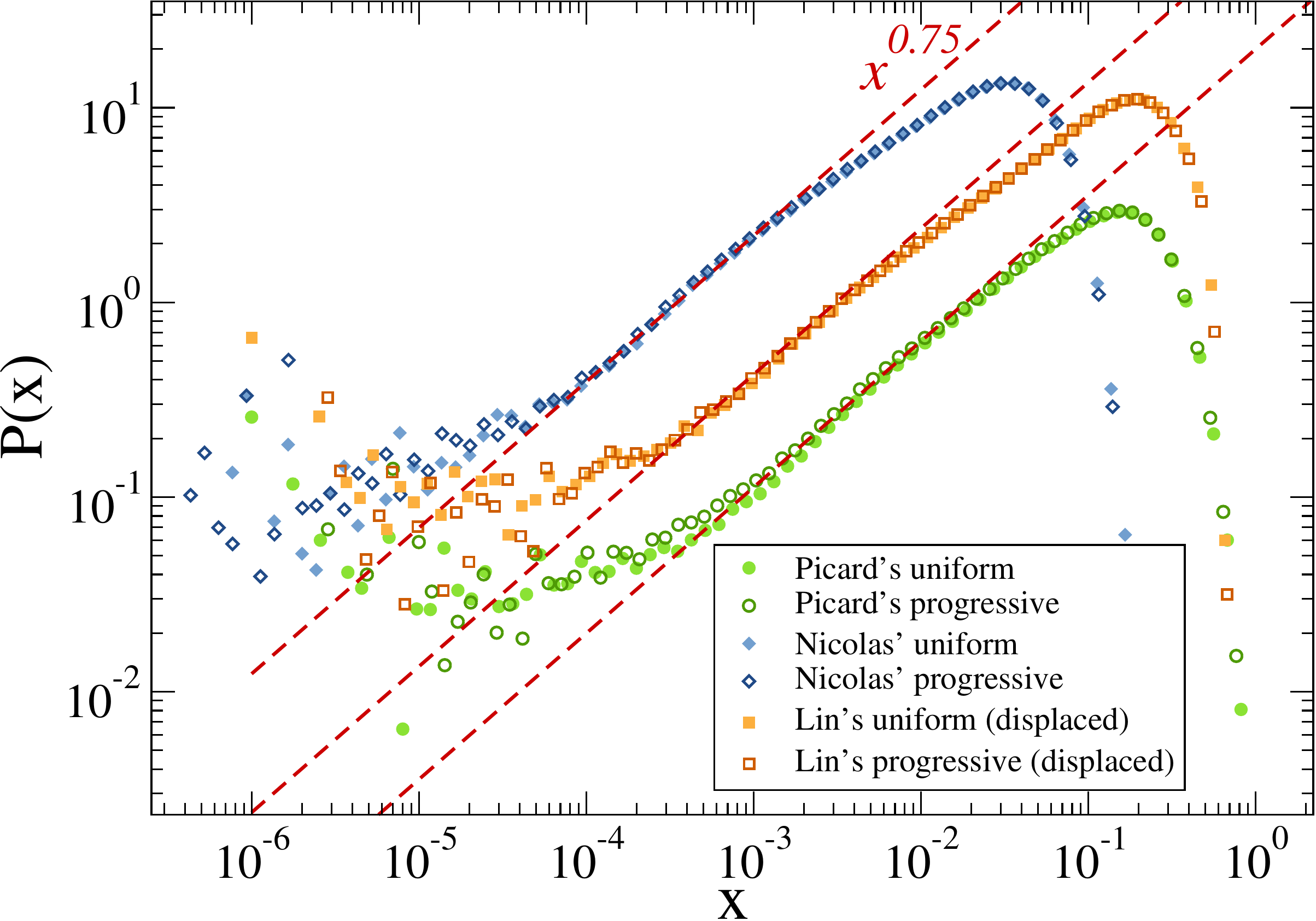} 
\end{center}
\caption{
$P(x)$ for different model variants as indicated in the labels and 
Red dashed lines display $\sim x^{\theta}$ with $\theta \simeq 0.75$
to guide the eye.
System size is $N=1024^2$ in all cases.
}\label{fig:PofX}
\end{figure}

In Fig. \ref{fig:PofX} we plot $P(x)$ for different EP models with $L=1024$.
Again, the values of $\theta$ that one could obtain from such curves vary
between $0.56-0.77$ according to the criterion used to identify the power-law regime.
Yet, in any case, the extracted exponent value is independent on each activation
rate used, as the correspondent distributions for the same model lay one on
top of the other.
Beyond the numerical uncertainty, the values of $\theta$ observed are
clearly above the value of $\vartheta=0.5$ extracted from the scaling of $\langle x_{\tt min} \rangle$.
Furthermore, when lines proportional to $x^\theta$ with $\theta=0.75$
are proposed, all model variants show a good compatibility with them
in a given range of $x$ (see Fig.\ref{fig:PofX}).
%
%This value of $\theta$ is consistent with the prediction of $\theta=\mu/2$
%in~\cite{lin2016mean}, considering $\mu=\beta$, as we discuss in Sec.\ref{sec:rae}.
\newtext{This value of $\theta$ happens to be consistent with the exponents found for
the flowcurve, as we discuss in Sec.\ref{sec:rae}. }

The same global form of $P(x)$, with an apparent plateau at $x\to 0$,
has been independently and simultaneously reported by Tyukodi
\textit{et al.}~\cite{tyukodi2019avalanches} and later confirmed in
MD simulations~\cite{ruscher2019residual} while this work was under review.
We further discuss this in Sec.~\ref{sec:PofXplateau}.

\subsection{Accumulated mechanical noise}
\label{sec:mechnoise}

%Although it has been reported in the literature that some critical exponents \newtext{of the yielding transition}
%are dimension-dependent, the long-range nature of the Eshelby interaction kernel
%makes us believe that the yielding transition would support an effective
%mean-field description, as soon as we correctly choose the distribution
%for a plugged mechanical noise in each finite dimension.
%
Standing on an arbitrary site $i$ in the system we can define 
$\xi_i(t)$ as the accumulated mechanical noise on $i$ at time $t$
due to avalanches occurring elsewhere. %, but not including site $i$.
We would like to study the statistical properties of this noise and
relate it to the observed values of the avalanche statistics and
the rheological exponent $\beta$.
In particular, we are interested in describing $\xi(t)$ as a correlated
noise with statistical properties defined by the so-called Hurst exponent $H$,
so that $\xi(x)$ is a (stochastic) homogeneous function of degree $H$,
\begin{equation}\label{eq:noiseHurst}
\xi(k x) \sim k^H \xi(x)
\end{equation}
Note that a standard random walk has $H=1/2$.
In general, in a mean-field description \cite{lin2016mean}, $H = 1/\mu$
where each contribution $\delta\xi$ to the accumulated noise $\xi$
is assumed to come from a distribution with a long tail 
\begin{equation}\label{eq:localNoiseLin}
P(\delta\xi) \sim \frac{1}{|\delta\xi|^{\mu+1}}
\end{equation}
and with $1\le \mu\le 2$.

\begin{figure}[!b]
\begin{center}
\includegraphics[width=1\columnwidth]{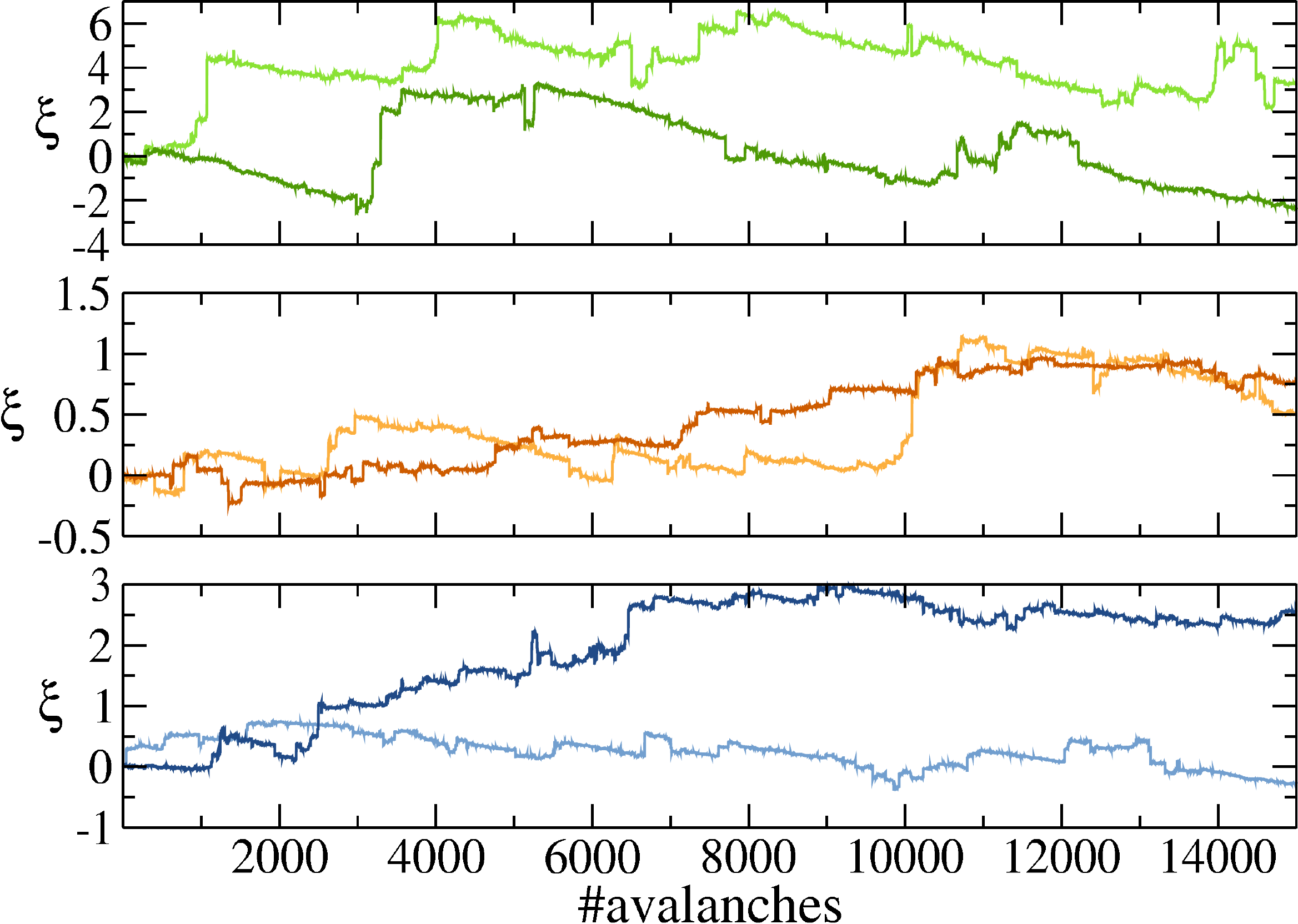} 
\end{center}
\caption{
Different realizations of the accumulated mechanical noise $\xi$ 
as a function of a quasistatic ``time'' represented by the number of avalanches
for Picard's (top panel), Lin's (mid-panel) and Nicolas' model  (bottom panel)
with both uniform (light colors) and progressive (dark colors) rates, and $L=256$.
}\label{fig:accumNoiseAR_realizations}
\end{figure}

Taking into account the action of individual plastic events through the
Eshelby propagator, \citet{lin2016mean} arrived to the conclusion that
$\mu=1$ is the only value with ``physical'' sense.
Nevertheless, if we discretize time in such a way that we can see the occurrence
of avalanches as the elementary event contributing to the mechanical noise, 
other values of $\mu$ may acquire a physical sense. 
In fact, a distant site at a given moment in the relevant time scale does
not feel the mechanical kick produced by a single site, but by a full avalanche
of sites yielding.
A coarse-graining in time is done by the system itself, which does not necessary
constraint the plastic events to occur one at a time and well separated
but overlaps them in a sudden burst of activity.
In a quasistatic loading protocol, the time scale separation 
between the duration of one avalanche and the loading time to trigger the next one
allows us to make this approximation and set the time-scale in avalanche counts.

Figure \ref{fig:accumNoiseAR_realizations} displays different realizations
of $\xi$ measured for various of our EPMs.
We can already notice with the bare eye the characteristic long steps or
L\`evy flights that distinguish this kind of signal from standard random-walks.
More importantly, we will show that in all cases these signals show
values of $H$ that are very close to each other, pointing to a robust
value of $H$ that defines an intrinsic property of the system.
To compute $H$ we can analyze the signal in the following way: for each
window of size $\epsilon$ in the avalanche sequence compute the
average value of the absolute noise difference in that window
$\left<\delta\right>(\epsilon)=\frac{1}{M}\sum_i^M |\xi(i)-\xi(i+\epsilon)|$.
$H$ is then the exponent of the relation $\left<\delta\right> \sim \epsilon^H $.
%The fitted exponent is sensitive to the range of $\epsilon$ where we fit.
%Discarding very small windows, where the noise is still basically dominated
%by a few avalanches, and large values of $\epsilon$ where the average on
%finite sizes lead us to a Brownian-like noise, we can fit a reasonable power
%law in a couple of decades, given values of $H$ in the range $0.67$-$0.69$
This is basically what the Detrended Fluctuation Analysis (DFA) (Nolds~\cite{nolds})
does after having subtracted from the signal a global trend.
Therefore, we collect large series of the accumulated mechanical noise at different
points in the system and analyze these signals with the DFA algorithm. 
\begin{figure}[!tb]
\begin{center}
\includegraphics[width=1\columnwidth]{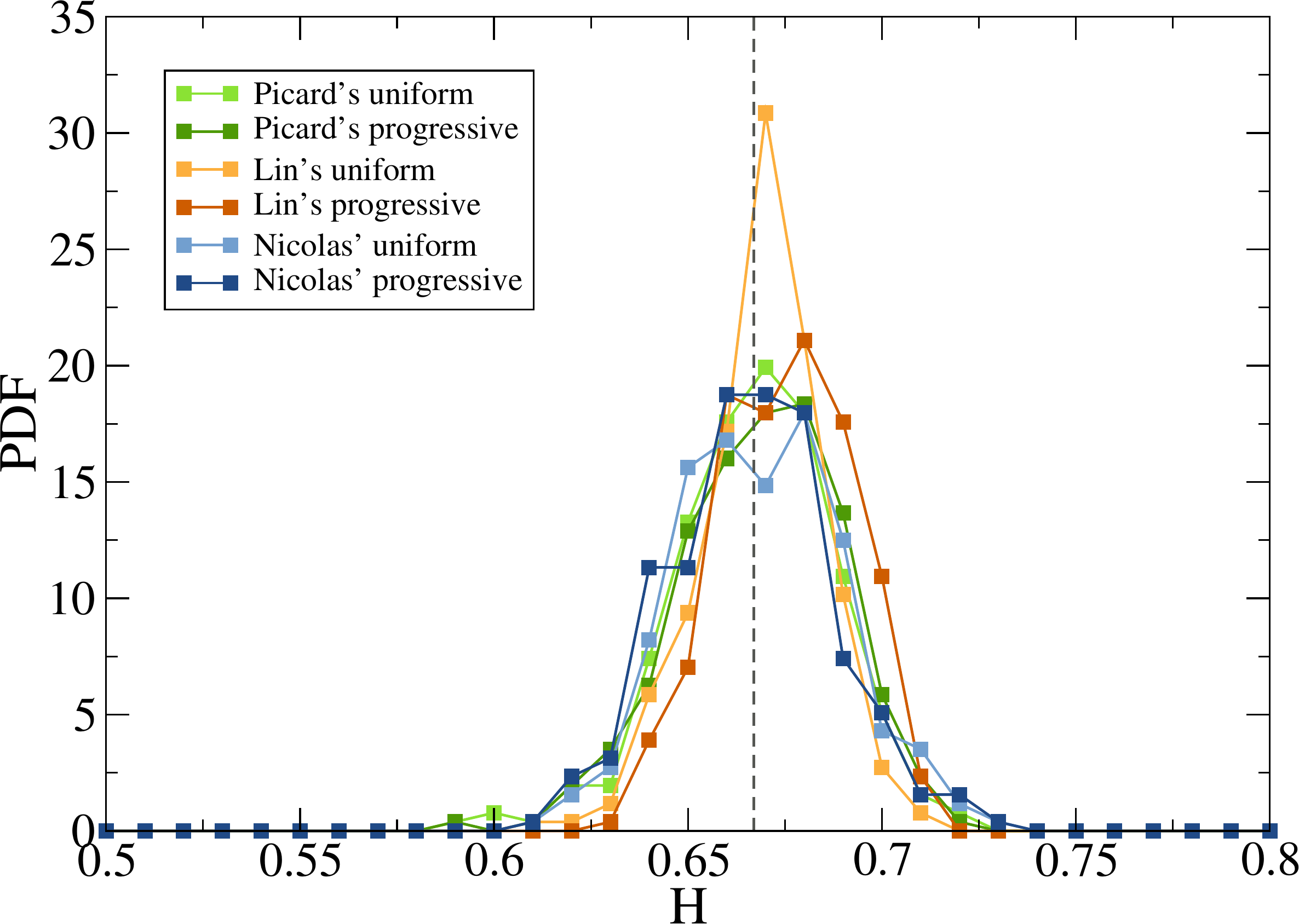} 
\end{center}
\caption{
The Hurst exponent analysis of all EPMs models in their two variants.
Histograms are built from the $H$ values obtained from accumulated noise 
signals taken at 256 different points in systems of size $N=256^2$.
The vertical dashed line corresponds to $H = \frac23$.
}\label{fig:hurstHistogram}
\end{figure}
Results for the histograms of $H$ values observed are shown in Fig.~\ref{fig:hurstHistogram}
for the different models and rate rule variants.
We observe that all systems show some spread around a mean value $\sim 0.67$,
but already the concurrence of all model variants around such a particular
value is non-trivial.
The DFA analysis is sensitive to details as the total signal length and the
minimum segment length for polynomial fitting, installing a non negligible
uncertainty in its output.
We believe that a good theoretical proxy for the Hurst exponent in 2d-EPMs 
is $H \simeq \frac23$, signaled by a vertical dashed line in Fig.~\ref{fig:hurstHistogram}.
This corresponds to $\mu=\frac{1}{H}\simeq 1.5$ in Eq.~\ref{eq:localNoiseLin},
subscribing the idea that a coarse-graining in time can give
a physical meaning to values of $\mu$ different from $\mu=1$.
The Hurst exponent is independent of the rate rule, what puts it at
the level of other static critical exponents such as $\tau_S$, $\theta$ or $d_f$.
In the Section \ref{sec:rae} below we provide arguments to believe
that the exponent $H$ characterizing the mechanical noise in a quasistatic
measure is useful for the estimation of the $\beta$ exponent of the flowcurve
as it departs from the critical yielding point.

%\begin{figure}[!tb]
%\begin{center}
%\includegraphics[width=1\columnwidth]{Figures/ARhurstsanalysis-crop.pdf} 
%\end{center}
%\caption{
%The hurst exponent analysis of Lin's model with uniform rates
%}\label{fig:ARhurst}
%\end{figure}

\subsection{Avalanche durations}

We now consider a critical exponent for which a dependence on the rate law may be expected.
This is the dynamical exponent $z$ that, combined with $d_f$, relates the avalanche size
with the avalanche duration.
The avalanche duration $T$ is expected to scale as a power law of the avalanche size $S$,
namely $T \sim S^{1/\delta}$.
If we assume that both $S$ and $T$ are controlled by a correlation length $\ell$ through
$S\sim \ell^{d_f}$ and $T\sim \ell^{z}$, then $T \sim S^{z/d_f}$ and $\delta = d_f/z$.

\begin{figure}[bt]
\begin{center}
\includegraphics[width=0.95\columnwidth]{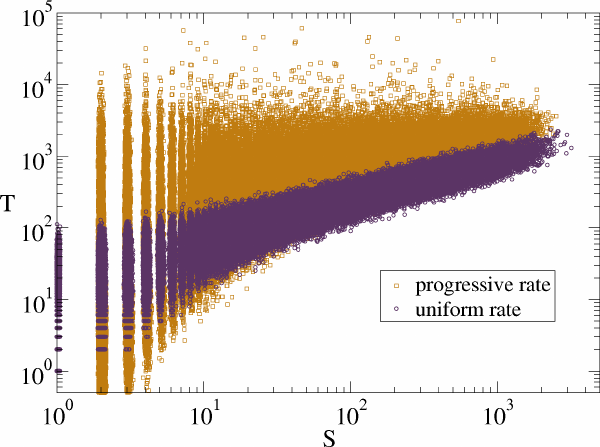}\\
\vspace{0.2cm} 
\includegraphics[width=1\columnwidth]{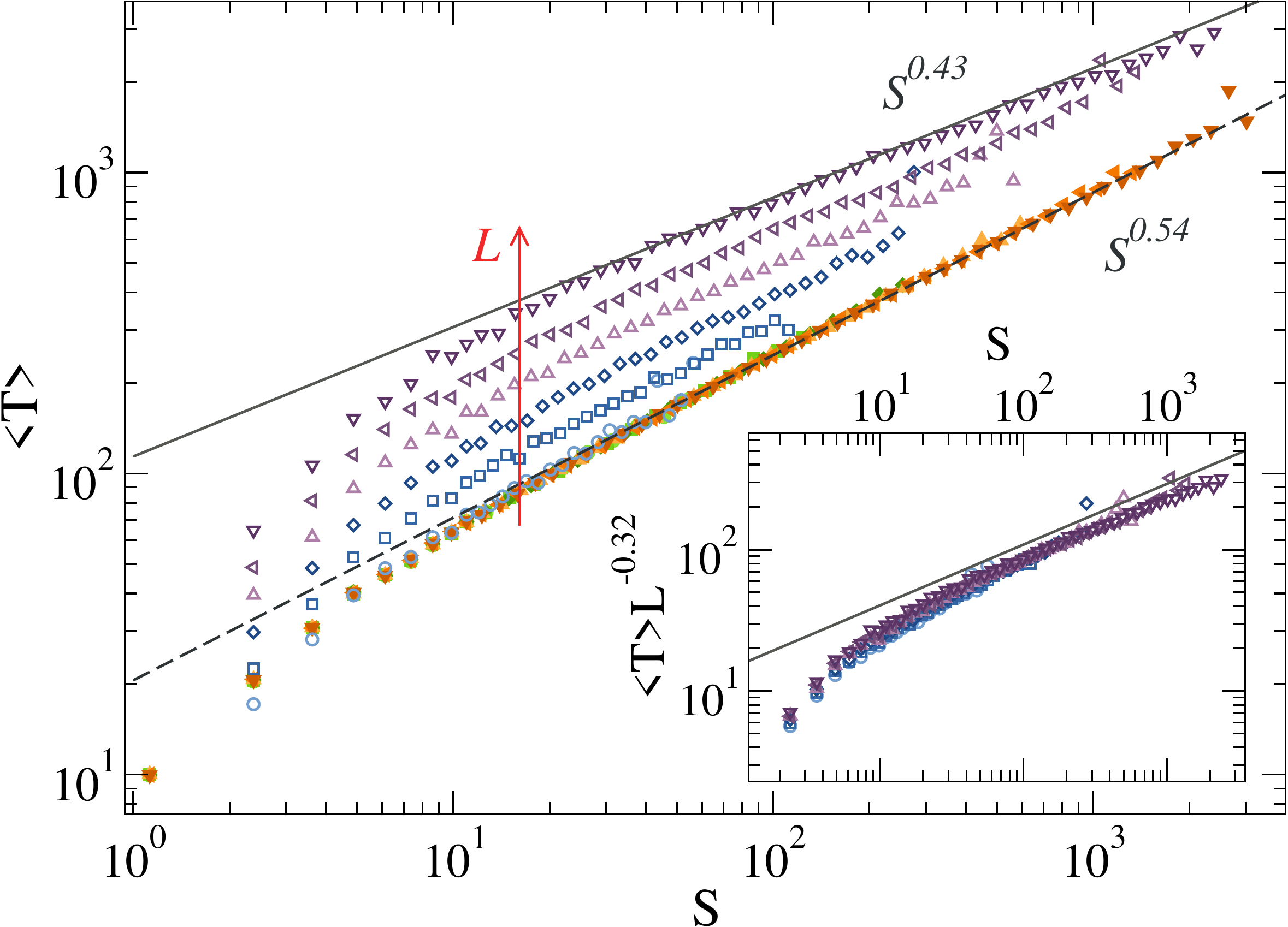} 
\end{center}
\caption{
\textit{Top panel:} Scatter plot of avalanche duration $T$ vs size $S$ for Lin's model with uniform and progressive rates
(linear system size $L=1024$).
\textit{Bottom panel:} Averaged avalanche duration as a function of the avalanche size for Lin's model with uniform and
progressive rates and different linear system sizes $L=32,64,128,256,512,1024$. 
The inset shows the scaling proposed in Eq. \ref{eq:TNS} for the progressive case, and derived in Appendix~\ref{ap:mf_duration}.
}\label{fig:TvsS}
\end{figure}

Fig. \ref{fig:TvsS}(top) shows the $T$-$S$ relation for a large set of avalanches
of quasistatic simulations at a fix system size ($L=1024$) from Lin's model with
both constant and progressive rates.
We can observe for both cases a broad cloud of points, 
clearly different for the two different rate rules.
Although scaling relations are difficult to guess from these clouds, it is clear that 
the progressive rate allows for a much larger spread of avalanche durations for
the same avalanche size, in particular, it allows for avalanches much longer in time.
Now, averaging the values of $T$ within small $S$ intervals, for different system sizes
we obtain the curves in Fig. \ref{fig:TvsS}(bottom).
For uniform rates, the results show a consistent power law with an exponent $z/d_f \simeq 0.54$.
Data for all system sizes overlap on the same curve.
Using the previously determined value $d_f \simeq 1.08$, we can obtain $z \simeq 0.58$. 
For progressive rates, the results are definitely different.
At a fix system size, a grow of $\left<T\right>$ vs $S$ with exponent $z/d_f \simeq 0.43$
is established for large enough avalanches, yielding for the dynamical exponent $z\simeq 0.46$,
i.e., lower than the uniform case.
Moreover, an additional feature is observed.
The results for increasingly large system sizes do not simply extend the region in which
a power law is observed but also produce a shift of the average duration towards
larger values.
 
This unexpected result can be rationalized in the following way.
Each time an avalanche triggers a new plastic event, its life is
expanded and some time will be added to its final duration.
For uniform activation rates, this time is on average $dT=\lambda^{-1}=1$.
On the other hand, for progressive rates, the time added to the avalanche duration
depends on the stress excess of the new yielding site;
on average $dT=\lambda^{-1} = (\sigma_i - \sigmaY_i)^{-\frac12}$.
The events that most contribute to the avalanche duration then, are the ones that
occur very close above their threshold. 
While their individual probability of yielding is low, the observation of these
events increases with the number of sites susceptible of being in that situation;
this is, it increases with increasing system size.
%
%In fact, this phenomenon of increasing average duration for a given avalanche size
%as system size increases suggests an alternative definition for the dynamical
%exponent, let's call it $\tilde z$,
%which relates the maximum avalanche size with the maximum time duration, 
%namely $T_{max}\sim S_{max}^{\tilde z/d_f}$.
%This dependence (sketched by the red dash-dotted line in Fig. \ref{fig:TvsS}(bottom))
%provides in the present case $\tilde z/d_f\sim 0.75$.
%
This previously unnoticed phenomenon is also present at a mean-field level, 
where a more quantitative estimation of the value of $z$ can be given and 
an exact scaling law derived (see Appendix~\ref{ap:mf_duration}),
leading to the $N$ and $S$ dependence of the avalanche duration:
\begin{equation}\label{eq:TNS}
T\sim N^\alpha S^{(1-\alpha)/2}
\end{equation}
We have tested this scaling in the inset of Fig. \ref{fig:TvsS}(bottom),
and it works very well with $\alpha \simeq 0.16$.
Notice that in the numerical determination of the durations $T$ reported in this section,
the time needed to destabilize the first site in the avalanche was not considered.
This was done in this way since, as the first site is destabilized by an infinitesimal
quantity, it would add a diverging contribution to the total time when progressive rates
are at play, completely spoiling the duration estimation.
For uniform rates the first site only adds an additional time unit, but for consistency
we do not considered it either in this case.

\subsection{Flowcurves}
\label{sec:flowcurves}

The results of the previous section about the dependence of $z$ 
on the activation protocol is not surprising as $z$ is the prototypical
dynamical exponent of the model.
Much more unexpected is the fact that the flowcurve exponent $\beta$ is
also dependent on the activation rate law, as we will discuss now.
To build flowcurves we use the strain rate controlled simulation protocol.
%
%Starting from a mechanically stable configuration (e.g. $\sigma_i=0, ~\forall i$)
We first deform with a large strain rate (e.g. $\gdot \simeq 0.5$) until a global
steady stress value is reached, and then average the measured stress in time.
After a measurement at a given strain rate, the strain rate value is reduced and the protocol is repeated,
covering a proper range in strain rates to draw the flowcurve in log-scale.

When arriving from the ``fluid'' side the approach to the yielding point is
continuous and the transition is critical.
As the strain rate $\gdot$ vanishes, the average stress reaches a finite value
that we call the (dynamical) yield stress $\sigma_c$.
In the vicinity of the limit $\gdot\to 0$ the flowcurve obtained by
a strain rate control protocol
$\left<\sigma(\gdot)\right> = \sigma_c + A\gdot^n$ can be also written as 
\begin{equation}
\gdot \propto (\left<\sigma(\gdot)\right> - \sigma_c)^\beta,
\end{equation}
where $\beta=1/n$.

%We use a finite $\gdot$ strain rate protocol to build the flowcurve.
%Starting from the state left by a higher $\gdot$, we deform a total strain
%of $\gamma_{\tt tr}=3$ to attain the steady state, and then deform
%further $\gamma_{\tt run}=10$ during which we take measurements each
%$\Delta\gamma_{\tt data}=0.05$.

\begin{figure}[!h]
\begin{center}
\includegraphics[width=1\columnwidth]{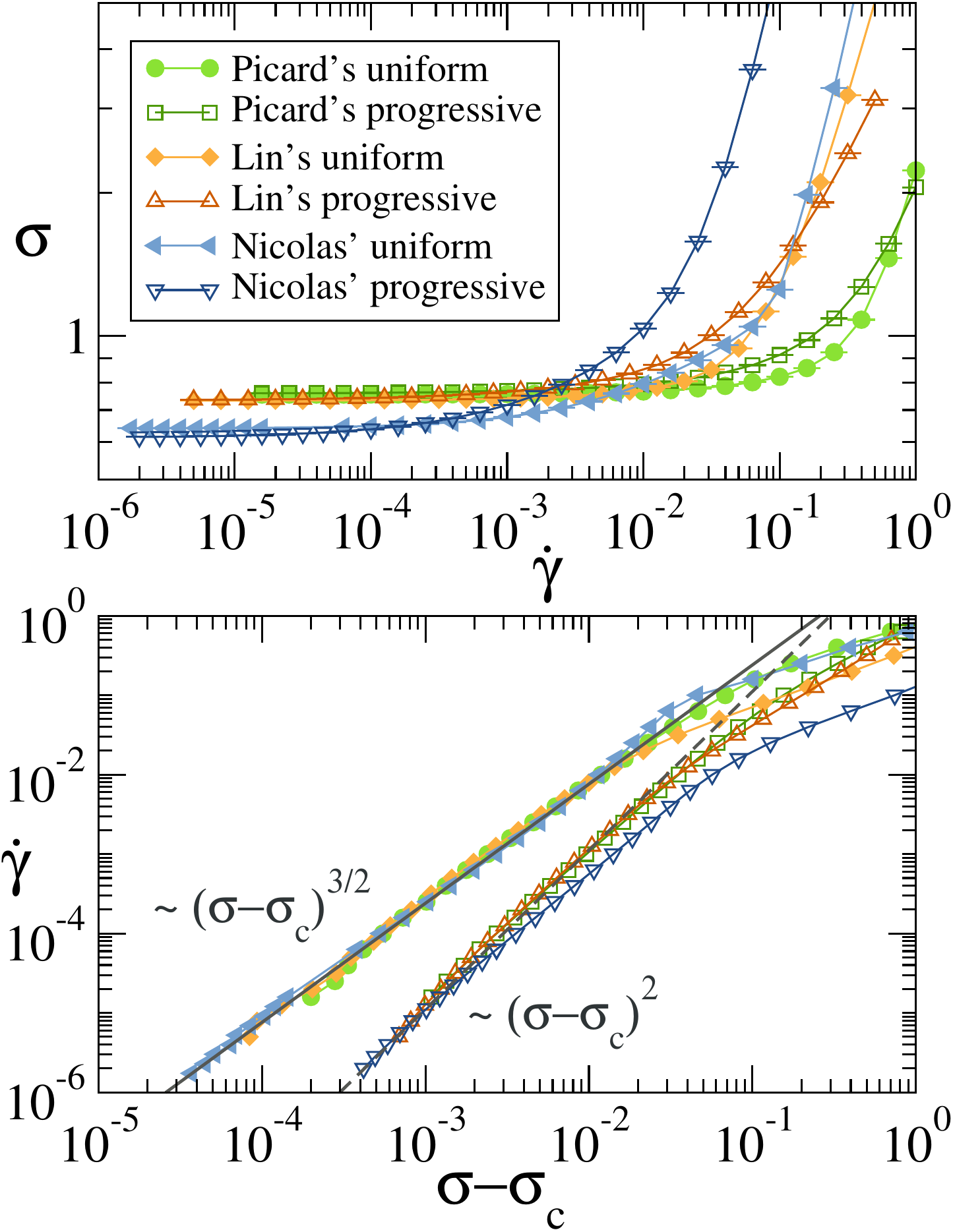} 
\end{center}
\caption{
Flowcurves for different elastoplastic models. 
Data corresponding to each model variant is identified by symbols and
colors as in the legend.
System size is $N=2048^2$.
\textit{Upper panel:} stress $\sigma$ vs. strain rate $\dot{\gamma}$.
\textit{Lower panel:} strain rate $\dot{\gamma}$ vs. stress excess $\sigma-\sigma_c$
(curves shifted arbitrary in the vertical direction for comparison).
Straight lines corresponding to power-law exponents $\beta = 3/2$ and
$\beta = 2$ are also displayed.
}\label{fig:flowcurves}
\end{figure}

Fig. \ref{fig:flowcurves} displays flowcurves for all three EPMs
with both constant and progressive transition rates for local yielding.
While the upper panel displays average stress vs. strain rate, in 
the lower panel we plot the strain rate as a function of the stress
excess above the dynamical yield stress.
One important detail in the construction of this curves is the appropriate
estimation of $\sigma_c$, for which we have implemented separately 
quasistatic measurements and averaged the stress in this limit over
long deformation windows.
Although it is a delicate fitting procedure, not free from
finite-size effects, % and very sensitive to the determination of $\sigmaY$
once $\sigma_c$ has been properly determined, all EPMs with uniform
rate show a flowcurve exponent compatible with $\beta = 3/2$.
as for the models with piece-wise parabolic
disorder potential analysed in ~\cite{FernandezAguirrePRE2018}.
Interestingly, when progressive rates are used instead, the exponent changes to $\beta = 2$
in all the three models taken from the literature\footnote{
The progressive case of the Nicolas's model in Fig. \ref{fig:flowcurves} 
shows a rather narrow range where $\beta=2$ seems to be valid.
We believe this is due to the relatively large time the variable $n_i$ can remain in
the "active" (i.e. $n_i=1$) state in this case, masking the effect of progressive
rates unless $\dot \gamma $ is very small},
suggesting that a local fluidization rate of the form $\lambda_{\tt prog} = (\sigma-\sigmaY)^\frac12$
takes us to the case of `smooth' disorder potentials~\cite{FernandezAguirrePRE2018} (see Section \ref{sec:rel_to_other_models}).
In any case, Fig.\ref{fig:flowcurves} clearly shows that the particular dynamical rule that
is used for local yielding has a direct impact in the value of the flowcurve exponent.

It can be noticed in passing that, by using progressive rates, the exponent
obtained in our 2$d$-systems is $\beta=2$, as the one derived for the mean-field
H\'ebraud-Lequeux (HL) model~\cite{Hebraud1998,AgoritsasEPJE2015,AgoritsasSM2017}.
Nevertheless, in the HL model the activation rate is constant.
The reason for this accidental coincidence rests in the fact that the flowcurve
exponent is not only determined by the transition rates, but also by the
mechanical noise statistics (which is dimension-dependent), as discussed
in the next Section.

\section{Discussion}
\label{sec:discussion}

\subsection{The residual density of sites at the verge of yielding}
\label{sec:PofXplateau}

In Sec.\ref{sec:xmin_and_Pofx} we have shown data for the distribution
$P(x)$ of $x=\sigmaY-\sigma$, the local distances to threshold,
showing that it develops a `plateau' as $x\to 0$.
This feature has been ignored or widely overlooked in the literature
so far, but has been put forward recently not only by a draft version
of this work but also in two other independent an  very recent preprints
by Tyukodi \textit{et al.}~\cite{tyukodi2019avalanches} (see also~\cite{tyokodi2016thesis})
and Ruscher\&Rottler~\cite{ruscher2019residual}.
Despite differences in the modeling approach, in particular the treatment of
periodic boundary conditions, our results and the results of
Ref.~\cite{tyukodi2019avalanches} are quite consistent.
We both observe for quasistatic simulations in the steady state 
that a plateau develops for the distributions $P(x)$ for different system sizes
at small $x$ and that the $\langle x_{\tt min}\rangle$ values lie between the
plateau and power-law regimes (see crosses in Fig.\ref{fig:PofXfinitesize}, an idea borrowed from Ref.~\cite{tyukodi2019avalanches}).
They observe a scaling of the plateau level $p_0\equiv P(x=0)$ with 
system size of the form $p_0\sim L^{-a}$ with $a=0.6$, while we report 
$a\simeq 0.55$ in Sec.\ref{sec:xmin_and_Pofx}.
Also, the $\phi$ exponents fitted are quite close, 1.35 and 1.33
in Ref.~\cite{tyukodi2019avalanches} and in our case, respectively.
The small discrepancy in the value proposed for $\theta$ (their 0.67 vs. our 0.75)
could be explained by the range of system sizes explored.
In MD simulations of 2D binary Lennard-Jones mixtures and using
a `frozen matrix method' the authors of Ref.~\cite{ruscher2019residual}
find a scaling analogous to the one of Fig.\ref{fig:PofXfinitesize}
with $a\simeq 0.65$ for the stationary regime (they also analyze the
MD `as-quenched' configurations prior to deformation), and a value of
$\theta\simeq 1.05$ even larger than ours.
They suggest the development of correlations in a preferential direction
at the microscopic level to justify a `depinning-like' form of $P(x\to 0)$.
We have explored several spatial configurations of $P(x)$ as-left after an
avalanche and we cannot hypothesize about the presence of correlations. 
Rather than thinking that way, we interpret the plateau effect at small $x$
as a consequence of working with a discrete time dynamics and a finite system size.
For a discrete time random walk in the presence of an absorbing border we know that
the probability amplitude right at the border is finite and its value decreases
with the amplitude of the time step.
Yet, we have performed simulations with time steps down to $dt=10^{-4}$ and the
modification of the obtained $\langle x_{\tt min}\rangle$ values is very slow.
The true origin of the finite size scaling for the $P(x)$ plateau at $x\to 0$
should have to do therefore not with the discretization in time of the individual
events evolution, but with the discreteness of the smallest kick felt by any site
in the system due to an avalanche happening elsewhere~\cite{inprep}.

It could be argued that being an effect that decreases with system size,
the plateau in $P(x)$ for small $x$ must eventually become irrelevant:
for sufficiently large system sizes the full $P(x)$ would acquire the expected
pure $P(x)\sim x^{\theta}$ form, and the scaling in Eq. (\ref{eq:xminfss})
(with $\theta$ instead of $\vartheta$) would hold.
But this is not so.
In Fig. \ref{fig:xminscaling} we see that $\langle x_{\tt min}\rangle$ is always
in between the power-law and the plateau region of $P(x)$, and then it is
mostly the scaling of this plateau with system size what determines the
value of $\vartheta$ and the scaling in Eq.\ref{eq:xminfss}.
Thinking the other way around, we do not find any strong reason why the value of
$\vartheta$ determined from Fig.~\ref{fig:xminscaling} should be equal to that detrmined from the
asymptotic form of $P(x)$ (the red dashed line in Fig. \ref{fig:PofXfinitesize}).
In fact, these values do not coincide as far as we can tell: 
while from Fig.~\ref{fig:xminscaling} we obtain $\vartheta \simeq 0.5$, the value from
the asymptotic form in Fig.~\ref{fig:PofXfinitesize} is rather $\theta\gtrsim 0.7$.
Actually, this situation is not new in the literature, where $\vartheta\sim 0.45-0.5$
has been determined from the scaling of $\langle x_{\tt min}\rangle$ but larger values
($\theta \simeq 0.6$) have been obtained from the full $P(x)$ in two-dimensional
systems~\cite{Karmakar2010,lin2014scaling,NicolasRMP2018}. 
%
%Besides a lack of a better understanding of the difference in the values
%of $\vartheta$ and $\theta$ determined by the two different methods, 
It should be always kept in mind that the exponent ruling the 
avalanche dynamics in the quasistatic limit is $\vartheta$,
namely the one obtained from $\left<x_{\tt min}\right>$,
since this is the average value of stress that is actually added to the
system after each avalanche to maintain it in a stationary state.
On the other hand, it is the value of $\theta$ (extracted from $P(x)$) the one
that one can relate to the flowcurve exponent~\cite{lin2016mean}, % (see also Sec. \ref{sec:rae}).
as we justify in the next section.

\subsection{Relations among exponents}
\label{sec:rae}

In the light of our results, we recall and discuss 
in this section some scaling laws proposed in the literature. 
To start with, an interesting relation among exponents comes from
a very simple argument
originally presented by~\citet{lin2014scaling}.
In the steady state, the stress has a well defined average and describes
a jerky plateau as a function of strain for any finite system size.
Therefore, on average, the stress increases in the loading phases
must be balanced by the stress drops during avalanches.
From the avalanches distributions $P(S)\sim S^{-\tau_S}f(S/L^{d_f})$, with $f(y)$
a fast decaying function for $y\gg 1$ and considering $S\equiv \Delta\sigma L^d$,
one obtains $\langle \Delta \sigma \rangle \propto L^{d_f (2-\tau_S)-d}$. 
On the other hand, the stress increment needed to trigger a new avalanche
is what we have called $x_{\tt min}$ which scales as 
$\langle x_{\tt min}\rangle  \propto L^{-\frac{d}{\vartheta+1}}$.
Equating $\langle \Delta\sigma \rangle \propto \langle  x_{\tt min}\rangle $, leads to
\begin{equation}\label{eq:scaling_tau_theta}
\tau_S=2-\frac{\vartheta}{\vartheta+1}\frac{d}{d_{f}}.
\end{equation}
By considering the values of $\vartheta\simeq 0.5$ (taken from
Fig. \ref{fig:xminscaling}), and $\tau_S\simeq 1.33$ and $d_f\simeq 1.08$
(from Fig. \ref{fig:PofS}), we verify that this relation fairly holds for our data.
This is not surprising since, as we explained, Eq. \ref{eq:scaling_tau_theta}
is originated in a simple requirement of stress stationarity in the system.
Note however that Eq.~\ref{eq:scaling_tau_theta} is not satisfied using
the $\theta$ value obtained from $P(x)\sim x^{\theta}$ in our data.

When the flowcurve exponent enter into the discussion, things are less clear.
Different relations have been proposed in the literature, in particular~\cite{lin2014scaling}
$\beta = 1 + \frac{z}{d-d_f}$ and~\cite{lin2017some} $\beta = 1 + \frac{1}{d-d_f}$,
that show either a weak agreement with simulations or an asymptotic agreement
in dimensions higher than $d=3$.
We start our discussion by considering the exponent $\mu$
governing the fat tails of the mechanical `kicks' $\delta\xi$
felt by a given site in the system, through the distribution~\cite{lin2016mean,lin2017some,FernandezAguirrePRE2018,jagla2018prandtl} 
\begin{equation}
P(\delta \xi) = \frac{A}{N} \left|\delta \xi\right|^{-\mu-1}
\end{equation}
with appropriate cutoffs.
When integrated in time, this noise produces a time signal that can be characterized
as a fractional Brownian motion with Hurst exponent $H$ that is given by 
\begin{equation}
H=\frac 1\mu
\end{equation}
In addition, it is also found analytically \cite{lin2016mean,lin2017some} that such
fractional Brownian motion generates a distribution $P(x)$ that behaves at low
$x$ as $x^{\theta}$, with 
\begin{equation}
\theta=\mu/2
\end{equation}
Lin \textit{et al.}~\cite{lin2016mean,lin2017some} have suggested that
$\mu$ is also the exponent $\beta$ of the flowcurve if it happens to be
larger than one.

Our numerical results, \newtext{together with} those in \cite{FernandezAguirrePRE2018,jagla2018prandtl} 
\newtext{(summarized in Section \ref{sec:rel_to_other_models}), strongly}
suggest that two dimensional yielding \newtext{should} be well described as a mean field transition
characterized by a value of $H=2/3$.
\newtext{In fact, that leads to a consistent value of $\theta=3/4$ as the one we measured
in Sec.\ref{sec:xmin_and_Pofx}.}
%which is compatible with our numerical results for this quantity.
%
%This value of $\theta$ is consistent with the prediction of $\theta=\mu/2$
%in~\cite{lin2016mean}, considering $\mu=\beta$, as we discuss in Sec.\ref{sec:rae}.
%
\so{As far as $\beta$ is concerned, results in \cite{FernandezAguirrePRE2018,jagla2018prandtl}
(summarized in Section \ref{sec:rel_to_other_models}) strongly suggest that a mean field
treatment of the yielding transition may apply also to the present case of EPMs.}
\newtext{Furthermore,} this generalized mean field analysis relates $H$ (and therefore $\theta$)
to the flowcurve exponent $\beta$ through 
\begin{equation}\label{eq:thetamu2}
\frac{1}{H} =  \beta - 1 + \frac{1}{\omega}. 
\end{equation}
where, $\omega$ is related to the form of the transition rates:
$\omega=1$ for uniform rates and $\omega=2$ progressive rates.
Note that the prediction in Lin et al. \cite{lin2016mean,lin2017some} (i.e. $\mu=\beta$) is consistent
with this equation for the usual case of constant rates ($\omega=1$).
\so{Also note that} In this case
the equations above lead to $\theta = \beta/2$
\footnote{Notice that in a previous publication by one of us~\cite{FernandezAguirrePRE2018},
the relation $\theta=\frac{1}{H}-1$ was proposed.
However, the derivation of this relation was based on a wrong assumption, and therefore is not correct.}
, an extremely simple relation between the flowcurve exponent and the power exponent of the
density of shear transformations, \newtext{that we verify in our data (see Sec. \ref{sec:flowcurves}).}
\newtext{It} makes perfect sense: %in the steady state in the quasistatic limit
The steeper the $P(x)$, the sooner many sites will yield when we set the system in motion,
the faster strain rate will increase and (equivalently) the slower the global stress will
increase with $\dot{\gamma}$ due to the frequency of local fluidization.
Notice that such a simple expression is also valid for the H\'ebraud-Lequeux model ($\beta=2$, $\theta=1$).

\begin{table}[!bt]
\caption{Values of the $\beta$ exponent according to Eq. \ref{eq:beta_H_relation0}
for different models: Prandtl-Tomlinson (PT), H\'ebraud-Lequeux (HL) and the
universe of 2$d$ EPMs.
}\label{tab:beta}
 \centering
 \begin{tabular}{l|ccc}
 \hline
 \hline
 \textbf{Rate type}    & \textbf{PT} & \textbf{HL} & \textbf{2$d$-EPM} \tabularnewline 
 \mbox{ }              & (\textit{H=1}) & (\textit{H=1/2}) & (\textit{H=2/3}) \tabularnewline 
 \hline
 Uniform               & 1           & 2           & 3/2               \tabularnewline 
 \hline      
 Progressive           & 3/2         & 5/2         & 2                 \tabularnewline 
%  \cline{2-5}
 \hline      
 \hline      
 \end{tabular}
\end{table}

The values of $\beta$ according to formula (\ref{eq:thetamu2}) for
different values of $H$ are summarized in Table \ref{tab:beta}.
The case $H=1$ corresponds essentially to the absence of stochastic noise, 
and this is realized in the standard one-particle Prandtl-Tomlinson model.
In this case the values of $\beta$ are well known. 
The value $H=2/3$ was argued before to correspond to the 2$d$-EPM, 
and we observe in fact that the values on Table~\ref{tab:beta}
are exactly the flow exponents that were obtained from the numerical
results in Fig. \ref{fig:flowcurves}.
The case of Gaussian noise $H=1/2$ with uniform rates is known to
correspond to the H\'ebraud-Lequeux model.  
The value $\beta=2$ obtained from Eq. \ref{eq:thetamu2} in fact
constitutes a well-known result for this model ($n=1/\beta=0.5$)~\cite{Hebraud1998,AgoritsasEPJE2015,AgoritsasSM2017}.

To complete the verification of the values contained in Table \ref{tab:beta},
which is to say, the accuracy of Eq.~\ref{eq:thetamu2}, it remains to
consider the HL model for progressive rates.
An analytical derivation of $\beta=5/2$ for that case is presented with some
detail in the %next Section
Appendix (\ref{ap:progressiveHL}).
In brief, by extending the standard rate of plastic events
$\nu_{\tt HL}(\sigma , \sigma_c) \equiv \frac{1}{\tau_S}\Theta(\sigma - \sigma_c)$
to
$\nu (\sigma,\sigma_c) \equiv \frac{1}{\tau_S}(\sigma-\sigma_c)^\eta \Theta(\sigma - \sigma_c)$,
so allowing for a progressive transition rate depending on the local stress excess,
we follow the derivation of the flowcurve ending up with
\begin{equation}
 \dot{\gamma} \sim (\sigma-\sigma_c)^{\eta+2}
\end{equation}
for general $\eta$. 
The case corresponding to smooth potentials has $\eta=\frac12$, obtaining $\beta=5/2$.
%
%So, allowing ourselves to do this analogy and taking $H=0.66$, we obtain
%$\beta_{\tt unif} \simeq 1.5$, $\beta_{\tt prog} \simeq 2.0$, which are
%exactly the flow exponents that we find in Fig. \ref{fig:flowcurves}.
%
In Table \ref{table:all} we summarize all the values of the exponents
we have obtained in the numerical simulations of the present work.

\begin{table}[!bt]
\caption{Values of critical exponents determined from numerical
simulations in two-dimensional systems.
Exponents in general do not depend on the particular model analyzed.
$\beta$ and $z$ differ for uniform and progressive rate rules.
}\label{table:all}
 \centering
 \begin{tabular}{l|cc}
 \hline
 \hline
 & \textbf{uniform} & \textbf{progressive} \\ 
% &  rate &  rate \\ 
 \hline
$\beta$ &  $1.5\pm 0.1$ & $2.0\pm 0.1$ \\ 
$\vartheta$ (from $\left(x_{\tt min}\right)$) & $0.5\pm 0.05$ & $0.5\pm 0.05$ \\ 
$\theta$ (from $P(x)$) & $0.75\pm 0.07$ & $0.75\pm 0.07$ \\ 
$\tau_S$  & $1.33\pm 0.03$ & $1.33\pm 0.03$ \\ 
$d_f$& $1.08\pm 0.05$ & $1.08\pm 0.05$ \\ 
$H$  & $0.67\pm 0.03$ & $0.67\pm 0.03$ \\ 
$z/d_f$  & $0.54\pm 0.02$ & $0.43\pm 0.04$ \\ 
%$\tilde z/d_f$  & undefined & $0.62\pm 0.05$ \\ 
$z$  & $\sim 0.583$ & $\sim 0.464$ \\ 
\hline
\hline
%$\tau_S ^*\equiv 2-\frac{\vartheta}{\vartheta+1}\frac{d}{d_f}$  & $1.38\pm 0.06$ & $1.38\pm 0.06$ \\ 
%$\beta ^*\equiv 1+\frac{z}{d-d_f}$ & $ 1.70 \pm 0.25$ & $1.47\pm 0.25$ \\ 
%$\tilde \beta ^*\equiv 1+\frac{\tilde z}{d-d_f}$ &  & $1.84\pm 0.25$ \\ 
% \hline
\end{tabular}
\end{table}

%%%%%%%%%%%%%%%%%%%%%%%%%%%%%%%%%%%%%%%%%%%%%%%%%%%%%%%%%%%%%%%%%%%%%%%%%%%

\subsection{Brief digression about the dynamical critical exponent $z$}

%To conclude this section, let us include a brief digression about 
%dynamical critical exponents.
%
%%%Following the discussion in the avalanches duration section...
%
Since the instantaneous long-range Eshelby interaction is responsible for the
dependence on microscopic properties and ultimately, for the different values
of $\beta$ and $z$ found, one can argue (as~\citet{lin2017some}) that spurious
effects will disappear when this `unphysical' instantaneous interaction is eliminated
(by adding a mechanism for propagating waves, for example).
While this might be the case, the use of an instantaneous interaction is not necessarily
an unphysical assumption that should be improved, but rather makes sense in many different
physical systems when restricted to the appropriate spatial and temporal scales.
In the depinning of magnetic structures for instance, long-range interactions
(typically dipolar) are considered, although these interactions cannot travel faster than
the speed of light! %\EF{[REF], quiza es mejor en este contexto ejemplificar con fracture propagation depinning Ponson et al.}
Also, the problem of contact line depinning is known to have a long-range elastic
interaction decaying as $1/r^2$.
This interaction is generated through the elastic propagation of interactions in the bulk
of the material, and in the end its propagation velocity must be
limited to the speed of sound in the material.
Nevertheless, that finiteness for the elastic propagation velocity has not been a major
obstacle for the theory there. %\EF{[REF]} 
In our case, an Eshelby instantaneous interaction is an approximation that considers the speed of sound
in the system to be infinite.
This is clear for instance in the analysis of~\citet{cao2018soft}, where in order to derive the Eshelby interaction
an infinite bulk modulus --and thus an infinite sound velocity-- is considered.
This approximation produces some results that are certainly not correct in a more realistic situation.
For instance a value of the dynamical exponent $z$ lower than one indicates a propagation of perturbations
as $r=C t^{1/z}$, and thus at velocities larger that any finite value if considering sufficiently large
times or distances.
The consideration of a finite sound velocity $c_s$ must provide ultimately a value of $z\ge 1$.
Yet the values we find for $z$ and $\beta$ and, furthermore their different values for different dynamical rules,
still make sense if we are analyzing cases (avalanche durations and sizes, for instance) for which $r/t=C t^{1/z-1}=C^zr^{1-z}\ll c_s$.
%It needs to be emphasized also, that the same argumentation about the limitations of a model with instantaneous 
%propagation as the one we use, can be made about the other many similar cases found in the literature, for which $z<1$ has been found.

%%%%%%%%%%%%%%%%%%%%%%%%%%%%%%%%%%%%%%%%%%%%%%%%%%%%%%%%%%%%%%%
%%%%%%%%%%%%%%%%%%%%%%%%%%%%%%%%%%%%%%%%%%%%%%%%%%%%%%%%%%%%%%%%%%%%%%%%%%%
\subsection{Relation to potential-energy-landscape-based models of yielding and justification of progressive rates}
\label{sec:rel_to_other_models}

A different class of mesoscopic models to study the yielding transition \so{exists,
which is}, based on a continuous description of the strain in the system
and the explicit inclusion of a quenched disorder landscape, \newtext{was proposed}~\cite{FernandezAguirrePRE2018}.
A brief analysis of \so{this kind of models} \newtext{them} will allow to clarify the meaning of 
uniform and progressive rates (Eq.\ref{eq:progressive_rate}) we have used in EPMs,
and \so{also} the origin of Eq. \ref{eq:thetamu2}.
These models can be written \so{in general} through \newtext{an} equation giving the time
evolution of the local strain in the system, $e_i$, in the form
\begin{equation}
\dot{e}_i = -dV_i/ e_i + \sum_j G_{ij}e_j + \sigma .
\label{eq:aguirre}
\end{equation}
where $G_{ij}$ is the same elastic propagator used in Eq. \ref{eq:eqofmotion1},
$\sigma$ is the applied stress, and $V_i$ indicate a set on random pinning
potentials on which $e_i$ evolve.

\so{On one side,} A generalized mean-field treatment for this kind of models was proposed
in~\cite{jagla2017non,FernandezAguirrePRE2018,jagla2018prandtl}. %(see also \ref{eq:eqofmotion1}),
\so{that} \newtext{It} accounts to replace the mechanical noise term in \ref{eq:aguirre} by a stochastic term
$\eta_i(t)$ with correlation characterized by its Hurst exponent $H$, leading to decoupled
equations of the form
\begin{equation}
\dot{e}_i = -dV_i/ e_i + \eta_i(t) + \sigma .
\label{eq:aguirre2}
\end{equation}
A detailed analysis of this stochastic one-particle model\cite{jagla2018prandtl} 
\so{leads to the conclusion} \newtext{shows} that the value of the flowcurve exponent
$\beta$ can be calculated once the values of $H$ and \so{some characteristics} \newtext{the analytical form}
of the pinning potentials \newtext{at the transition point between potential wells} are know.
The main \so{final} result is:
\begin{equation}\label{eq:beta_H_relation0}
\beta  =1 + \frac{1}{H} - \frac{1}{\omega}.
\end{equation}
where \so{$\omega$ is related to the form of the pinning potentials at the transition point
between different potential wells:}
$\omega=2$ for smooth potentials, and $\omega=1$ for potential with discontinuous forces.
\newtext{We} now \so{we need to} link this with EPMs.

\begin{figure}[!tb]
\begin{center}
\includegraphics[width=.75\columnwidth]{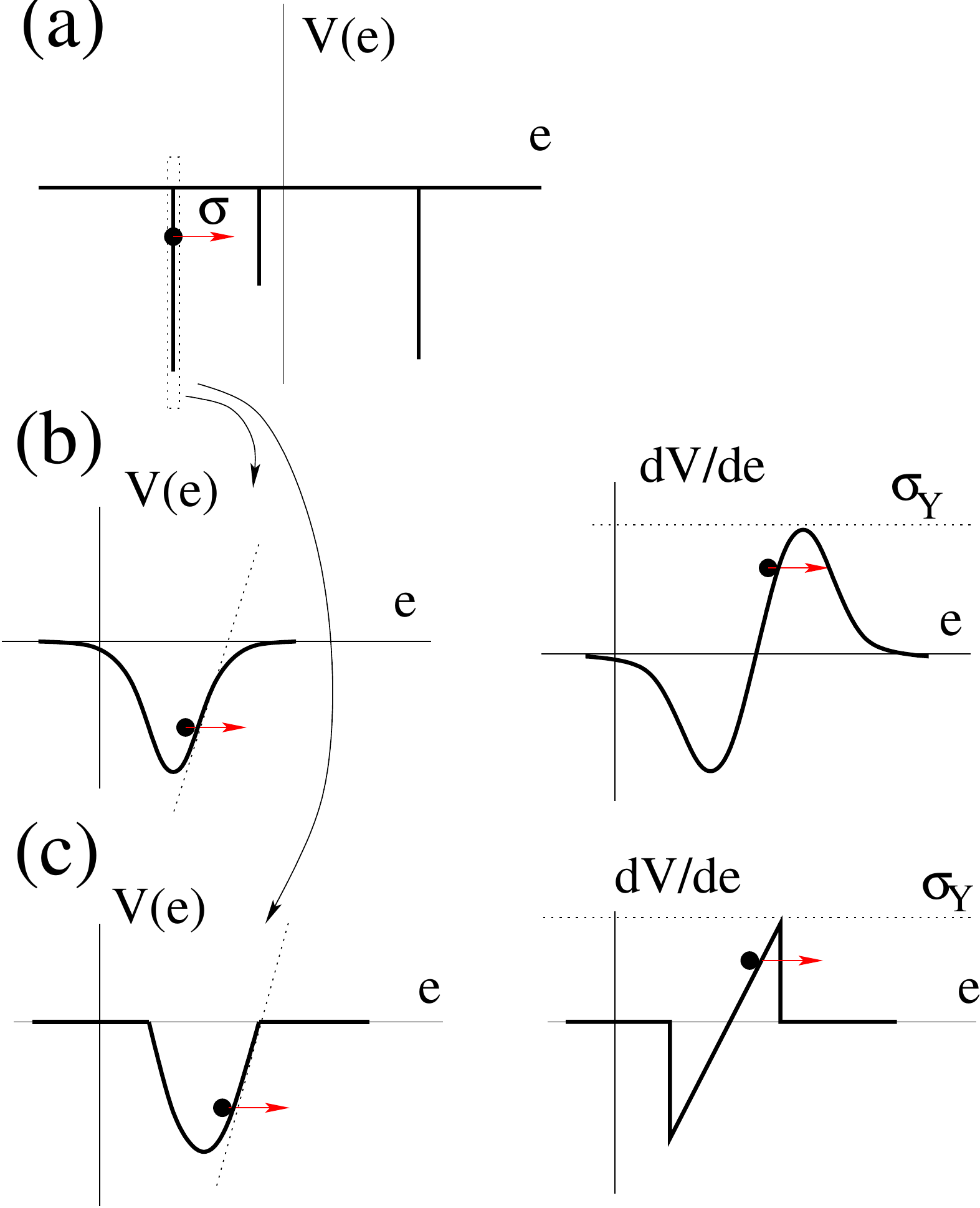} 
\end{center}
\caption{(a) Sketch of the pinning potential acting on the strain variable $e_i$
in the narrow well approximation of a model for yielding as considered in Ref~\cite{FernandezAguirrePRE2018}.
$e_i$ will jump to the next well when the applied stress overpasses the local yield
stress $\sigmaY$.
Although narrow, the form of the well remains important in determining the dynamics
that the site follows upon this jumping.
(b) Smooth \so{pinning case} \newtext{depinning}.
(c) Sharp \so{pinning centers} \newtext{depinning}.
}
\label{app1}
\end{figure}
Notice that all EPMs we have studied in this paper share the same kind of dynamics for
the local stress $\sigma_i$:
When $\sigma_i$ overpasses the local yield stress $\sigmaY$, 
it enters a stage in which there is a probability per unit time that the site becomes plastic. 
The average time for this stochastic 
activation  process is note the ``activation time'' $\tau_{\tt on}$.
The inverse of $\tau_{\tt on}$ time is the ``activation rate'' $\lambda$.
Once in the plastic state ($n_i=1$) the site evolves
towards a new relaxed position, decreasing its stress and accumulating plastic strain.
The time related with this second stage of the local yielding event varies enormously
with the \newtext{particular EP} model.
\so{:
in \cite{lin2014scaling} it is instantaneous,
in \cite{Picard2005} it is stochastic with rate $\tau_{\tt off}$ (=1 in our study),
in \cite{nicolas2014rheology} it depends on the strain accumulated locally.
}
Yet, it is expected \so{, in particular when progressive rates for the activation
are introduced,} that the activation time will dominate the total duration of
the plastic event.

The same kind of behavior for the local stress is \so{also} realized in models like the
one of Eq. \ref{eq:aguirre}.
\so{To see its relation with EPMs clearly it is convenient to} 
Consider the limiting case of a `narrow wells potential' $V$.
In this case, the local strain $e_i$ evolves in a random disordered pinning potential
as sketched in Fig. \ref{app1}(a).
The wells are assumed to be extremely narrow, in such a way that the value of $e_i$
is constant within the well~\footnote{In the analogous EPM, $n_i=0$ while the site
is in the well and $n_i=1$ when it is out.}.  
Each well is characterized by the force that it has to be applied in order for the
site to escape from it.
This threshold force is the analogous of the local yield stress $\sigmaY$ in EPMs.
As long as the locally applied stress $\sigma$ (originated in the externally applied stress
and the contribution from all other sites through $G_{ij}$) is lower than $\sigmaY$,
the value $e_i$ remains constant.
When $\sigma>\sigmaY$ the value of $e_i$ increases, and eventually reaches the value
corresponding to the next potential well.
Due to the increase of $e_i$ there is also a decrease of the local stress, similarly
to what happens in EPMs.
The important point to consider is the time $\tau_e$ it takes for this process to occur,
i.e., for $e_i$ passing from a given potential well to the next one.
To evaluate $\tau_e$, its is not enough to know that potential wells are narrow, we must
consider also their shape.
Referring to Fig. \ref{app1}(b), \newtext{let's} suppose \newtext{that} we have some smooth
pinning well defined by some potential energy function.
\so{Under this condition,}
The maximum of the derivative of this potential will define the local yield stress.
If the overstress $\delta \sigma \equiv \sigma_i-\sigmaY$ is small, the time to escape
from the well $\tau_e$ can be estimated by solving an equation of the form
\begin{equation}
\dot e_i \sim -e_i^2-\delta\sigma
\label{eq1}
\end{equation}
with initial condition $e_i\sim 0$ at $t=0$, and looking for the time at which $e_i\sim 1$.
A direct solution of the equation, or \newtext{a} dimensional analysis \cite{jagla2018prandtl},
provides $\tau_e \simeq \delta\sigma^{-1/2}$.
This is the reason to consider --in the model analyzed in this paper-- a ``progressive''
transition rate of the form $\lambda_{prog}\sim \delta \sigma^{1/2}$.

\so{Note that} 
The previous argument can be generalized to potential wells with a more
general form at the escape point, \so{(that of the maximum of its derivative)} 
like $|e|^\psi$, instead of $e^2$ in Eq. \ref{eq1}.
The analysis of this case yields $\lambda_{prog}\sim \sigma^{(\psi-1)/\psi}$.
However, to have potential wells with this characteristic requires an extremely fine tuning
of the potential, that is required to be non-analytical at the point of its maximum derivative.
The only sensible possibility (in addition to the smooth case) is that of a potential that
sharply vanishes at a finite value of $e$ (Fig. \ref{app1}(c)).
This case corresponds in fact to $\psi=1$ and provides $\lambda=cte$
\so{. In this way we justify}
\newtext{; and so, justifies} our assertion that constant activation rates can be assimilated to pinning
potentials with successive wells separated by singular points in which there is a jump in the
pinning force.
\so{Finally,}
Given this equivalence, Eq. \ref{eq:beta_H_relation0} can \so{also} be considered to be valid
\newtext{also} for EPMs, identifying the $\omega=1$ case with constant transition rates, and $\omega=2$ with
progressive transition rates.

%%%%%%%%%%%%%%%%%%%%%%%%%%%%%%%%%%%%%%%%%%%%%%%%%%%%%%%%%%%%%%%%%%%%%%%%%%%%%%%%
\section{Summary}
\label{sec:summary}

In this paper we have investigated different versions of elasto-plastic models (EPMs)
discussed in the literature addressing their critical properties at the yielding transition.
Accurate numerical simulations have revealed that for the three cases analyzed all
critical exponents are the same, thus suggesting that these values are universal, independent of model
details.
Nevertheless, we have identified a dynamical rule in EPMs on which ``dynamical'' critical exponents
depend upon.
This is the form of the rate law used to fluidize sites that have exceeded the local yielding threshold.
Previous implementations of EPMs in the literature considered this rate $\lambda$ to be a constant,
independent of the degree of overstress above the threshold value.
We have gone beyond that, analyzing also the case of a progressive rate that depends on the degree
of overstress, in particular $\lambda \sim (\sigma_i-\sigmaY_i)^{1/2}$.
This change of dynamical rule has a strong effect on the flowcurve exponent
$\beta$ (which changes from $\beta=3/2$ to $\beta=2$) and the dynamical exponent $z$.
The change of rule impacts in exactly the same manner on the three models analyzed.
Furthermore, we have investigated also the effect of progressive rate in mean-field.
For the well known H\'ebraud-Lequeux model the inclusion of progressive rates 
transforms the flowcurve exponent from the standard $\beta=2$ value to
$\beta=5/2$ in the progressive case.
Critical exponents that we have called ``static'' do not depend on the yielding rate rule:
in particular, the exponents $\tau_S$ and $d_f$ for the avalanche size distribution and
the exponents describing the density of sites at the verge of yielding.
For the latter, we find a distribution of the form $P(x)\simeq P(0) + x^\theta$.
Therefore, $\vartheta=d/\phi-1$ extracted from the finite-size scaling of
$\langle x_{\tt min} \rangle \sim N^{-\phi}$, happens to be different from $\theta$.
Both are independent on model and rate-rule details.
Also the Hurst exponent $H$ derived from the accumulated mechanical noise
signal is a static and model-independent exponent.
It further allows to propose an alternative mean-field approximation to yielding
which we find to give good predictions for the exponents extracted from simulations
of fully spatial EP models in this work.

Conceptually, we believe that progressive yielding rates (i.e., 
smooth pinning potentials) should be more representative of real systems. 
It would be more difficult to think of a real system as having an effective sharp potential.
Even though a direct comparison with experiments could be difficult, molecular dynamics
simulations can be set to deal with interaction potentials of the smooth or sharp type,
and investigate if the same variation of $\beta$ we find here %, on the kind of potential we have found here
is also found in those atomistic simulations.
%The exponent obtained for this case ($\beta=2 \Rightarrow n=0.5$) is, in fact, closer
%to the typical value (within broad dispersion) found in the laboratory~\cite{BonnRMP2017}.

The finding of the critical exponent $\beta$ depending on
microscopic details of the disorder potential for the yielding transition has arisen
somehow as a surprise, since it is commonly expected (based in the evidence provided
by renormalization group arguments for the depinning transition) that these microscopic
details should be irrelevant.
A second thought, however, shows that this influence of dynamical rules that we
observe in yielding is analogous to the one observed in depinning in the mean-field
fully-connected case, where $\beta$ also depends on the particular shape of the disordered
potential (see for instance~\cite{KoltonPRE2018}).
Actually, the elastic coupling $G_{\bf q}$ for yielding (Eq. \ref{eshelby_kernel}) 
is in fact independent of the absolute value of ${\bf q}$ (exactly as in the fully
interacting depinning case, where $G_{\bf q} = -A$, with $A$ a constant for all $\bf q$),
generating a long range interaction in real space $G(r)\sim 1/r^d$.
This `sorority' between the yielding transition of amorphous solids %analyzed from EPMs
and the fully-connected mean-field depinning classical problem is subject of analysis
of a separate work~\cite{FerreroJaglaMF}.

\section*{Conflicts of interest}
There are no conflicts to declare.

\section*{Acknowledgements}
EEF acknowledges support from grant PICT 2017-1202, ANPCyT (Argentina).

\appendix

\section{Appendix}

\subsection{Progressive rates in the H\'ebraud-Lequeux model}
\label{ap:progressiveHL}

The H\'ebraud-Lequeux model~\cite{Hebraud1998} is defined by the evolution
of the probability distribution function ${\mathcal{P}(\sigma,t)}$ of 
local stress $\sigma$ at time $t$, under an external strain rate ${\dot{\gamma}(t)}$ as
\begin{equation}
\label{eq:HLequation}
\begin{split}
 \partial_t \mathcal{P}(\sigma ,t)
 =	&	- G_0 \dot{\gamma}(t) \partial_\sigma \mathcal{P}
 		+ D_{\tt pl}(t) \partial_{\sigma}^2 \mathcal{P} \\
 	& 	- \nu (\sigma , \sigma_c) \mathcal{P} + \Gamma (t) \delta (\sigma)
\end{split}
\end{equation}
where the rate $\nu$ is given by
\begin{equation}\label{eq:HLrate}
\nu_{\tt HL}(\sigma , \sigma_c) \equiv \frac{1}{\tau}\Theta(\sigma - \sigma_c)
\end{equation}
(with $\Theta$ the Heaviside function and $\delta$ the Dirac distribution)
i.e., the rate of plastic events is approximated by a fixed value $1/\tau$ in any overstressed region,
and the plastic activity is defined from
\begin{equation}
 \Gamma(t) = \frac{1}{\tau} \int_{\sigma'> \sigma_c} d \sigma' \mathcal{P}(\sigma' ,t),
\end{equation} 
Finally, a key ingredient of the model is that the diffusion coefficient $D_{\tt pl}(t)$
is assumed to be proportional to the plastic activity:
\begin{equation}
\label{eq:DHL}
 D_{\tt pl}(t) = \tilde{\alpha} \Gamma(t)
\end{equation}
where $\tilde{\alpha} >0$ is an ad-hoc parameter of the model.
In a stationary situation, as the plastic activity is proportional to $\dot\gamma$
for low shear rates~\cite{AgoritsasSM2017}, the previous equation also implies that $\dot\gamma\sim  D_{\tt pl}$.
For any $\tilde{\alpha}$ smaller than a critical value $\tilde{\alpha}_c=\frac{\sigma_c^2}{2}$,
a non-trivial solution exist for the (nonlinear) evolution in ${\mathcal{P}(\sigma,t)}$
%the nonlinearity being encoded in the diffusion coefficient and leading to the non-trivial features of the HL model.
that translates in a flowcurve ($\left<\sigma\right> = \sigma_c + A \dot{\gamma}^n$) with $n=1/2$ (i.e., $\beta=2$).

In the following we will show that modifying Eq.(\ref{eq:HLrate}) to 
\begin{equation}\label{eq:HLrate2}
\nu (\sigma,\sigma_c) \equiv \frac{1}{\tau}(\sigma-\sigma_c)^\eta \Theta(\sigma - \sigma_c)
\end{equation}
i.e., introducing a stress-dependent progressive rate in the HL model,
the flowcurve results in
\begin{equation}
\left<\sigma\right> = \sigma_c + A \dot{\gamma}^\frac{1}{2+\eta}
\end{equation}

Rather than attempting a full solution of Eq.(\ref{eq:HLequation}) with the modified 
rate (\ref{eq:HLrate2}), we simply concentrate in tracking the impact of the functional
form of the local plastic rate in the scaling properties of the stress close to the
critical point.
First, notice that the result $\dot{\gamma}\sim D_{\tt pl}$ at first order~\cite{AgoritsasEPJE2015,AgoritsasSM2017}
is still valid in this case.
At the same time, the stress excess $\Delta\sigma \equiv \sigma - \sigma_c$
when we impose a small strain rate is proportional to the probability density at 
the local threshold $\mathcal{P}(\sigma_c)$.

Looking for a stationary solution of Eqs. (\ref{eq:HLequation},\ref{eq:HLrate2})
at zero order in $\dot{\gamma}$, we can write for $\sigma>\sigma_c$
\begin{equation}
0 = D \partial_{\sigma}^2 \mathcal{P} - (\sigma-\sigma_c)^\eta \mathcal{P}
\end{equation}
The solution for $\mathcal{P}$ should have the form
\begin{equation}
\mathcal{P} = \mathcal{P}(\sigma_c) f\left( \frac{\sigma-\sigma_c}{D^{\frac{1}{\eta+2}}}\right)
\end{equation}
with $f(0)=1$.
Now, asking for the usual closure equation of the HL model (\ref{eq:DHL})
that provides self-consistency, we obtain
\begin{eqnarray}
 D & = & \tilde{\alpha} \frac{1}{\tau} \int_{\sigma'> \sigma_c} d \sigma' (\sigma'-\sigma_c)^\eta \mathcal{P}(\sigma') \\
  & = &  \tilde{\alpha} \mathcal{P}(\sigma_c)  \int_{\sigma'> \sigma_c} d \sigma' (\sigma'-\sigma_c)^\eta f\left( \frac{\sigma-\sigma_c}{D^{\frac{1}{\eta+2}}}\right) \\
  & = &  \tilde{\alpha} \mathcal{P}(\sigma_c) D^{\frac{\eta+1}{\eta+2}} \int_{u_0} du ~u^\eta f(u)  
\end{eqnarray}
The last integral is just a number, then we conclude that
\begin{equation}
D \sim \mathcal{P}(\sigma_c) D^{\frac{\eta+1}{\eta+2}}
\end{equation}
and
\begin{equation}
\Delta\sigma \sim \mathcal{P}(\sigma_c) \sim D^{\frac{1}{\eta+2}} \sim \dot{\gamma}^{\frac{1}{\eta+2}}
\end{equation}
In other words, 
\begin{equation}
 \dot{\gamma} \sim (\sigma-\sigma_c)^{\eta+2}
\end{equation}
for the HL model modified with progressive rates for local yielding of the form $\sim(\sigma-\sigma_c)^\eta$.
In the particular case corresponding to smooth potentials we must take $\eta=\frac12$, obtaining $\beta=5/2$
which is precisely the value predicted by Eq. (\ref{eq:beta_H_relation0}) for this case.

\subsection{Avalanche duration vs. size in mean-field}
\label{ap:mf_duration}

A toy model with discrete pinning sites and stress dependent transition rates can be used
to investigate analytically the dependence between duration and spatial extent of avalanches.
% and then to obtained the asymptotic form (for large avalanches) of this behavior, and in particular the value of $z$ ala function of $\alpha$. 
%
Imagine a pinning potential consisting of narrow wells in which an elastic interface is pinned.
Each well has a threshold force value that is necessary to apply to unlock the interface from it.
Once a local threshold is exceeded the interface locally jumps to the next well with probability $\lambda$
(the inverse of the time it will take a particle to move from one well to the next).
$\lambda$ may be a constant or a quantity depending on the force excess over the threshold:
$\lambda=(f_i-f_i^{th})^\alpha$, where $f_i$ is the actual value of the force applied at site $i$,
$f_i^{th}$ is the local threshold force, and $\alpha$ is a numerical exponent.
A configuration of the system is characterized by the values of $f_i$, which are the elastic forces
at every site.
The configuration is stable if $f_i<f_i^{th}$.
On this configuration an avalanche is triggered by increasing uniformly the force up to the point
where $f_i$ becomes equal to $f_i^{th}$ at some particular $i$, which then becomes unstable. 
After some time the unstable site jumps to the next well, $f_i$ is reduced by some quantity $\delta$,
and all $f_j$ move up a quantity $\delta/N$.
This may produce some new sites to become unstable (i.e., to overpass their $f^{th}$).
The process continues until there are no more unstable sites.
At this point an avalanche of some duration $T$ and some spatial extent $S$ has occurred.

\begin{figure}
\includegraphics[width=0.8\columnwidth]{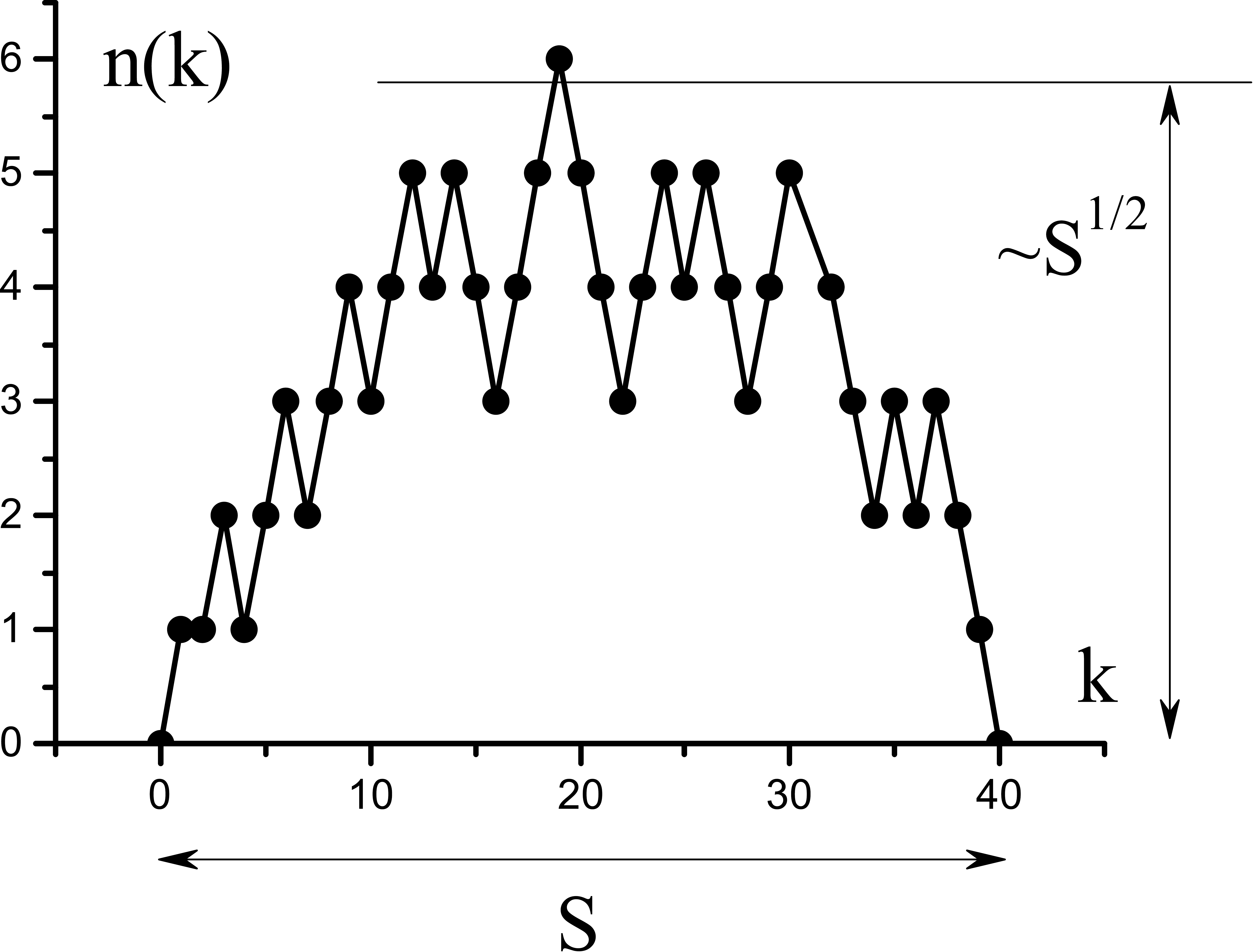}
\caption{Instantaneous size of the avalanche as a function of the number
of particles that already jumped to new equilibrium positions.
This plot is a random walk.
The duration $T$ of the avalanche can be evaluated once the times $dt_k$
to pass from $k$ to $k+1$ are known.
\label{esquema}
}
\end{figure}

We can obtain the relation between $T$ and $S$ for this avalanche as follows.
Suppose we plot the number of unstable sites $n$ as a function of the number of sites $k$
that have already jumped to their new positions within the avalanche.
Such a plot is a random walk.
The avalanche size $S$ is the number of sites that have jumped when there are no more
unstable sites, namely $n(S)=0$.

In order to calculate $T$ for a given $S$ we have to sum all time intervals $dt_k$
between successive values of $k$, namely
\begin{equation}
T=\sum_{k=1}^{S-1} dt_k
\end{equation}

Let us consider first the simplest case in which $\lambda$ is constant.
This means that any individual site jumps in a time that is always $\sim 1$.
If there are $n(k)$ unstable sites at some moment, the typical time for the
first one of them to jump is $dt_k\sim 1/n(k)$.
In addition, the average form of $n(k)$ for a random walk between two zero
crossings at 0 and $S$ is $n(k)\sim\sqrt{k(S-k)}/L^{1/2}$ then we obtain
\begin{equation}
T\sim\sum_{k=1}^{S-1} \frac{S^{1/2}}{\sqrt{k(S-k)}}
\label{2}
\end{equation}
that for sufficiently large avalanches can be written (using $x\equiv k/S$) as
\begin{equation}
T\sim\sum_{x=\frac 1S,\frac 2S,...,1-\frac 1S} \frac{S^{-1/2}}{\sqrt{x(1-x)}}=S^{1/2}\int_0^1 
\frac{dx}{\sqrt{x(1-x)}} \propto S^{1/2}
\label{3}
\end{equation}
i.e., we obtain $z=1/2$ which is the standard value of the dynamical exponent for
depinning in mean-field.

The previous calculation can be extended to the case of an arbitrary value of $\alpha$
in $\lambda=(f_i-f_i^{th})^\alpha$.
It requires the evaluation of $dt_k$ in this situation.
The calculation is now non-trivial, because every unstable site has a different value
of $y_i\equiv f_i-f^{th}$, and then a different transition rate.
The value of $dt_k$ can be calculated as
\begin{equation}
dt_k=\left( \sum_{i=1}^{n_k} y_i^\alpha \right )^{-1}
\label{4}
\end{equation}
(note that if $\alpha=0$, $dt_k=1/n(k)$ and we return to the previous case).
To calculate $dt_k$ the actual distribution of $y_i$ values if needed.
We focus on its calculation now.

Consider a situation with $n$ unstable sites with values $y_1$ ... $y_n$.
We will work in a continuum limit, with $n$ large.
We want to calculate the expected distribution of $y$, that will be characterized
by a function $P(y)$, such that 
\begin{equation}
\int_0^\infty P(y)=n. 
\label{int=n}
\end{equation}
As the rate goes as $y^\alpha$, we can write
\begin{equation}
\frac{dP}{dt}=Py^\alpha
\end{equation}
The equilibrium condition for $P(y)$ is that when one of the $y_i$ jumps
(and then disappears as an unstable site) and all the distribution is
shifted by $1/N$ the configuration remains stable.
This leads to
\begin{equation}
\frac{dP}{dt}=C_0\frac{dP}{dy}
\end{equation}
The constant $C_0$ remains yet undetermined, and will be fixed by normalization below.
Combined with the previous equation this gives
\begin{equation}
C_0\frac{dP}{dy}=Py^\alpha
\end{equation}
and from here
\begin{equation}
P(y)=P_0\exp{\left(-\frac{C_0y^{\alpha+1}}{\alpha+1} \right)}
\end{equation}
The two constants are determined from the normalization condition (Eq. \ref{int=n}), and from the fact that 
$P(0)=N$, since on average, a shift in $y$ of $1/N$ produces the appearance of one new unstable site.
The final result is 
\begin{equation}
P(y)=N\exp{\left[-a_0\left(\frac{Ny}{n} \right)^{\alpha+1}\right ]}
\end{equation}
with $a_0$ a numerical constant.
The continuous form of Eq. (\ref{4}) allows to calculate the average time
interval up to the next particle jump $dt_k$ as
\begin{equation}
dt_k=\left( \int_0^{\infty} P(y)y^\alpha dy\right)^{-1}
\end{equation}
Doing the integration, the result is
\begin{equation}
dt_k\sim \frac{N^\alpha}{n^{\alpha+1}}
\end{equation}
%Some values of the numerical constant $A_\alpha$ are: $A_0=...$, $A_{1/2}=...$, $A_1=...$. \EF{Values missing}
Using this expression to perform the same analysis that was done before in Eqs. (\ref{2}), (\ref{3})
leads to (assuming $\alpha < 1$)
\begin{equation}
T\sim\sum_{k=1}^S \frac{ N^\alpha S^{(\alpha+1)/2}}{(k(S-k))^{(\alpha+1)/2}}
\label{22}
\end{equation}
\begin{equation}
T\sim\sum_{x=\frac 1S,\frac 2S,\ldots,1-\frac 1S} \frac{N^\alpha S^{-(\alpha+1)/2}}{(x(1-x))^{(\alpha+1)/2}}\sim N^\alpha L^{(1-\alpha)/2}
\label{33}
\end{equation}
This is the final result.
It provides the value of the dynamical exponent as $z=(1-\alpha)/2$ and, at the same time, a dependence on the system size $N$.
The latter disappears in the uniform rate case ($\alpha=0$), corresponding to sharp pinning potentials.

%%%END OF MAIN TEXT%%%


\balance

\renewcommand\refname{References}

%%%REFERENCES%%%
%\bibliography{ReviewBibClean}
%\bibliographystyle{unsrt}
%\bibliographystyle{rsc} %the RSC's .bst file

\providecommand*{\mcitethebibliography}{\thebibliography}
\csname @ifundefined\endcsname{endmcitethebibliography}
{\let\endmcitethebibliography\endthebibliography}{}

\end{document}